\documentclass{tut2016_hide} % put ./tut2016 if using only locally
\usepackage[colorlinks=true,breaklinks=true,bookmarks=false,urlcolor=blue,citecolor=blue,linkcolor=blue,bookmarksopen=false,draft=false,hypertexnames=false]{hyperref}

\usepackage[english]{babel}
\usepackage[autostyle, english = american]{csquotes}
\MakeOuterQuote{"}

\usepackage{bbm,xspace,multirow,multicol}
\usepackage{cleveref}
\crefname{subsection}{subsection}{subsections}
\crefname{problem}{problem}{problems}
\crefname{assumption}{assumption}{assumptions}
\usepackage[normalem]{ulem}
\usepackage{algorithm,algpseudocode}

\newcommand{\eps}{\varepsilon}
\newcommand{\bI}{\mathbbm{1}}
\newcommand{\bE}{\mathbb{E}}

\newcommand{\cF}{\mathcal{F}}

\newcommand{\LP}{\mathsf{OPT}_\mathsf{LP}}
\newcommand{\Offer}{Y}
\newcommand{\Free}{\mathsf{Free}}
\newcommand{\vx}{\mathbf{x}}
\newcommand{\gRO}{\gamma^\mathsf{RO}}

\usepackage[square,numbers]{natbib}
\begin{document}
%%%%%%%%%%%%%%%%

\CHAPTERNO{}% Enter chapter number here, e.g. "Chapter 1"
\TITLE{Randomized Rounding Approaches to Online Allocation, Sequencing, and Matching}    % Enter chapter title here
%\SUBTITLE{}% Enter substitle (only if enecessary) and outcomment

% Enter author(s) here:
\AUBLOCK{% 
\AUTHOR{Will Ma}
\AFF{Graduate School of Business and Data Science Institute, Columbia University, New York, NY 10027, \EMAIL{wm2428@gsb.columbia.edu}}
} % This ends \AUBLOCK

% \CHAPTERHEAD{Ma: Sequential Randomized Rounding}% Shortened running head "Author(s): Short Title"

\ABSTRACT{%
Randomized rounding is a technique that was originally used to approximate hard offline discrete optimization problems from a mathematical programming relaxation.
Since then it has also been used to approximately solve sequential stochastic optimization problems, overcoming the curse of dimensionality.
To elaborate, one first writes a tractable linear programming relaxation that prescribes probabilities with which actions should be taken.
Rounding then designs a (randomized) online policy that approximately preserves all of these probabilities, with the challenge being that
the online policy faces hard constraints, whereas the prescribed probabilities only have to satisfy these constraints in expectation.
Moreover, unlike classical randomized rounding for offline problems, the online policy's actions unfold sequentially over time, interspersed by uncontrollable stochastic realizations that affect the feasibility of future actions.
This tutorial provides an introduction for using randomized rounding to design online policies, through four self-contained examples representing fundamental problems in the area: online contention resolution, stochastic probing, stochastic knapsack, and stochastic matching. \\
% These problems find application in online resource allocation, search, dynamic scheduling and learning, revenue management and e-commerce fulfillment, often providing a unifying abstraction for disparate domains.
% Compared to other approaches for these problems, randomized rounding is often simpler and faster, having approximation ratio guarantees that also illustrate the benefits of adaptivity or deferral, and having fairness implications that are not native to dynamic programming.

\noindent INFORMS 2024 Tutorial.
} % This ends the abstract

 % This ends the keywords

%%%%%%%%%%%
\maketitle  % This will make the chapter opening ("titlepage")
%%%%%%%%%%%

\section{Introduction}

In online decision-making, one must commit to a sequence of decisions while facing uncertainty about the future.
We consider online decision-making problems where we have full probabilistic knowledge of the way in which future uncertainties may unfold, and the performance of a decision-making policy is defined by expected reward.
In these problems, an optimal policy that maximizes performance can always be described using dynamic programming.
However, computing such a policy is often intractable, due to an exponential state space
in the dynamic programming.

We take instead a relax-and-round approach to these
problems.
In the first step, we write a computationally tractable Linear Program (LP) that is a \textit{relaxation}, which means that its optimal objective value $\LP$ is no worse than the performance of dynamic programming.
In the second step, we "round" a solution of this LP to construct an online decision-making policy in a computationally tractable manner.

We illustrate the Relax-and-Round approach through the following problem.
There are agents $i=1,\ldots,n$, each with a known weight $w_i\ge0$ and a known independent probability $p_i$ with which they will show up.
The policy observes sequentially whether each agent shows up, and must immediately decide accept or reject for each agent who shows up.
There is capacity to accept up to $k$ agents, and the performance of a policy (to be maximized) is the expected total weight of agents it accepts.

For the first step, we write the following LP relaxation:
\begin{subequations} \label{lp:prophet}
\begin{align}
\max\quad & \sum_{i=1}^n w_i x_i
\\ \mathrm{s.t.}\quad & \sum_{i=1}^n x_i \le k \label{lp:prophet:capacity}
\\ & 0\le x_i\le p_i &\forall i=1,\ldots,n \label{lp:prophet:showup}
\end{align}
\end{subequations}
This LP is computationally tractable because it is polynomially sized.
To see that it is a relaxation, suppose we set each variable $x_i$ equal to the probability of an optimal policy accepting agent $i$.
Constraint~\eqref{lp:prophet:capacity}, which says that at most $k$ agents are accepted in expectation, is satisfied because the policy accepts at most agents $k$ on every sample path.
Constraint~\eqref{lp:prophet:showup} is satisfied because the policy cannot accept an agent $i$ with probability greater than the probability $p_i$ with which they show up.
This proves that setting the $x_i$'s equal to the acceptance probabilities of an optimal policy defines a feasible solution to the LP, whose objective value $\sum_i w_i x_i$ equals the performance of the optimal policy.
Therefore, $\LP$ is no less than the performance the optimal policy; i.e., the LP is a relaxation.

For the second step, given a feasible solution $(x_i)_{i=1}^n$ to the LP, the goal is to find an online policy that accepts every agent $i$ with probability at least $cx_i$, for a constant $c$ as large as possible.
We call this a \textit{sequential randomized rounding (SRR)} problem.
To explain why, let $X_i\in\{0,1\}$ indicate whether our online policy accepts agent $i$.
The goal can be restated as to randomly "round" the fractional vector $(x_i)_{i=1}^n$ into an integer vector $(X_i)_{i=1}^n$, approximately preserving acceptance probabilities in that
\begin{align} \label{eqn:rounding}
\bE[X_i] &\ge c x_i &\forall i=1,\ldots,n.
\end{align}
Importantly, the rounding is sequential, where an online policy must decide each $X_i$ without knowing which future agents will show up, while biding by the problem constraints that
at most $k$ agents can be accepted
and
only agents who show up can be accepted.

Generally, that goal of SRR is to establish a constant $c\in[0,1]$ such that~\eqref{eqn:rounding} can be satisfied for any instance of the problem and any feasible solution to the LP, in a computationally tractable manner.
This would imply a \textit{$c$-approximate algorithm}, which is an algorithm that can compute tractably for any instance an online policy whose performance is at least $c$ times that of an optimal policy.
To see why, we can run the policy satisfying~\eqref{eqn:rounding} on an optimal solution $(x_i)_{i=1}^n$ to the LP, and the expected total weight of agents accepted would be at least $\sum_i w_i (cx_i)$, i.e.\ the performance would be at least $c\cdot\LP$.
This in turn is at least $c$ times the performance of an optimal policy, by virtue of the LP being a relaxation.
Constant $c$ is often referred to as the \textit{approximation ratio}.

That being said, the SRR approach has additional implications beyond approximation ratios.
In fact, for this problem a $c$-approximate algorithm is meaningless, because an optimal policy can be directly computed using dynamic programming (the state space is not exponential).
However, for this problem the SRR approach additionally implies that an online policy can have performance at least $c$ times that of the \textit{offline} policy, which knows in advance the agents who will show up and accepts from among them the $k$ agents with highest weights.
To see why, suppose we set each $x_i$ equal to the probability of the offline policy accepting agent $i$.  It follows from the same argument that constraints~\eqref{lp:prophet:capacity}--\eqref{lp:prophet:showup} are satisfied, and hence the performance $\sum_i w_i x_i$ of the offline policy can be feasibly attained in the LP.
Therefore, it is no greater than $\LP$, and the SRR policy that has performance at least $c\cdot\LP$ also has performance at least $c$ times that of the offline policy.
Turning to a second additional implication of SRR, we note that~\eqref{eqn:rounding} promises \textit{every} agent $i$ that they will be accepted with probability at least $c$ conditional on showing up.
This can be interpreted as a form of "fairness", which is not satisfied by dynamic programming.

\subsection{Outline of Tutorial}

This tutorial serves as an introduction to SRR, expounding on concepts and providing details that are difficult to find in academic papers.
It also provides a list of references to more advanced topics that can serve as a (brief) survey.

We describe four self-contained problems that represent fundamental applications of SRR.
The first is an SRR problem by definition, whereas SRR is used to derive state-of-the-art approximation algorthms for the latter three problems.
\begin{itemize}
\item \textbf{Online Bayesian Selection (\Cref{sec:selection})}: we study an SRR problem that is essentially identical to the one from the Introduction, motivated by a mobile pantry allocating food supplies while satisfying fairness across agents.
We discuss the differences between agents being encountered in a fixed order (\Cref{sec:selectionK}) vs.\ random order (\Cref{sec:selectionRandomOrder}), and explain how these are special cases of the Online Contention Resolution Scheme (OCRS) problem (\Cref{sec:ocrs}).
\item \textbf{Probing and Search (\Cref{sec:hiring})}: we study the simplest stochastic probing problem, motivated by a hiring firm that dynamically decides on the order to send offers to candidates.
We write an LP relaxation and solve the corresponding SRR problem to derive approximation ratios and adaptivity gaps.
We later extend to problems where the firm must first interview applicants to decide their value, as well as the ProbeMax problem (\Cref{sec:ptk}).
\item \textbf{Stochastic Knapsack (\Cref{sec:stocKnap})}: we study the stochastic knapsack problem, which can be viewed as a stochastic scheduling problem where rewards are collected for completing jobs in a finite horizon.
We write an LP relaxation and solve the corresponding SRR problem to derive an approximation ratio, where the technique can be flexibly extended to finite-horizon Markovian bandits problems.
\item \textbf{Stochastic Matching (\Cref{sec:matching})}: we study variants of the maximum matching problem in graphs that have both sequential decisions and stochastic realizations.  We first apply the SRR result from \Cref{sec:selection} to derive a 1/2-approximate algorithm for online stochastic matching, before specializing to IID arrivals (\Cref{sec:knownIID}), discussing stochastic probing and search in graphs (\Cref{sec:stocProbMatching}), and returning to OCRS for the matching polytope (\Cref{sec:ocrsGraphs}).
Some of these problems were instrumental to the development of the SRR approach, but are deferred to the end due to being more advanced.
\end{itemize}
We conclude with a summary of the key techniques and concepts in SRR learned through this tutorial, along with a discussion of
future directions
% recent directions and open questions
(\textbf{\Cref{sec:conc}}).

% This tutorial by no means suggests that SRR is the only approach to sequential stochastic optimization.
% Instead, it is meant to provide a complementary perspective to the well-known approaches of (approximate) dynamic programming and reinforcement learning (\citet{bertsekas2012dynamic}), compared to which SRR can have the following virtues:
% \begin{itemize}
% \item Its randomized algorithms often run quickly, and have simple conceptual interpretations;
% \item Its algorithms come with worst-case approximation ratio guarantees;
% \item These guarantees characterize the benefits of higher-level changes, changing the order of decisions, and/or choosing this order adaptively;
% \item In the example above, SRR promises a uniform probability to every agent for being allocated a supply, providing a form of fairness that is not native to the notion of value-to-go functions in dynamic programming.
% \end{itemize}

% We mention some related tutorials and surveys covering topics outside our scope.

\subsection{Related Materials}

First, the online allocation and matching problems we study often consider the "competitive ratio" against the offline policy.
Positive results for SRR can generally be translated into positive results for competitive ratio, but we do not focus on this in the tutorial, instead referring to surveys about prophet inequalities (\citet{lucier2017economic,correa2019recent}) and online matching (\citet{mehta2013online,huang2023applications,huang2024online}).
We note that the online matching literature also considers competitive ratios in the "adversarial" setting, in which case one can use "online" randomized rounding procedures that do not know the inputs to be rounded in advance.
By contrast, we focus on "sequential" randomized rounding in stochastic settings where one does know the input probabilities in advance.

Second, we limit our scope to sequential randomized rounding for sequential stochastic optimization problems, whereas randomized rounding itself has a much longer history in approximating NP-hard integer programming problems in combinatorial optimization.  Classical examples of randomized rounding can be found in many chapters of books on approximation algorithms (\citet{vazirani2001approximation,williamson2011design}).

Finally, we restrict to Bayesian settings where the given probability distributions are assumed to be correct.  Meanwhile, several tutorials have been written about distributionally robust optimization (\citet{kuhn2019wasserstein,blanchet2021statistical}), where the goal is to hedge against distribution shifts.

\section{Online Bayesian Selection} \label{sec:selection}

We study a fundamental problem in sequential randomized rounding problem that we motivate using fairness.
We will later use it as a building block for online resource allocation in \Cref{sec:matching}.

\begin{problem}[$k$-unit Rationing] \label{prob:selection}
There is a set of agents $[n]$, where we will generally let $[n]$ denote $\{1,\ldots,n\}$. 
A mobile food pantry produces $k$ units of food supplies at the start of each day and drives by the agents in fixed order $1,\ldots,n$.
As long as food remains upon reaching an agent $i$, the mobile pantry can optionally offer a unit of food to the agent.
If the mobile pantry offers food, then the agent takes it independently with probability (w.p.) $x_i$, representing the probability of agent $i$ needing food on a given day that is known to the mobile pantry.
The mobile pantry aims to maximize the probability $\gamma$ with which it can promise every agent that they will be offered food.
The mobile pantry can skip offering food to earlier agents in order to ration food for later agents, but it cannot turn around to return to a skipped agent.
\end{problem}

\begin{remark}
An alternate, equivalent formulation of \Cref{prob:selection} is that the mobile pantry first asks if the agent needs food, and if so, decides whether to give food.
This alternate formulation is better known in the literature, as the Online Contention Resolution Scheme (OCRS) problem which has a wide array of applications (see \Cref{sec:ocrs}).
Regardless, in this tutorial we focus on the formulation in \Cref{prob:selection} instead, which originally appeared in \citet{alaei2011bayesian} as the "Magician's problem".
\end{remark}

An optimal rationing policy for \Cref{prob:selection} requires randomization.  For example, suppose there is $k=1$ food supply and $n=2$ agents with $x_1=x_2=1/2$.
If the mobile pantry always offers food to agent 1, then it would get taken with w.p.~$x_1$, leaving food for agent 2 w.p.\ only $1-x_1=1/2$.
On the other hand, if the mobile pantry always skips agent 1, then agent 1 is being offered food w.p.~0.
In this example the mobile pantry can in fact promise $\gamma=2/3$ to both agents: each day, it randomly decides to offer food to agent 1 w.p.~2/3; it then offers any remaining food to agent 2.  Agent 1 would be offered food w.p.~2/3, while agent 2 would be offered food w.p.
\begin{align*}
1-\Pr[\text{agent 1 was offered food}]\Pr[\text{agent 1 needs food}]=1-\frac23\cdot\frac12=\frac23.
\end{align*}
This is the optimal policy and promise guarantee for this example.

\citet{alaei2011bayesian} showed how to solve this rationing problem in general, when $k=1$.
Indeed, consider any fixed $\gamma>0$ that the mobile pantry would like to guarantee.
If there is food remaining when the mobile pantry drives by agent $i$, then it must offer food w.p.
\begin{align} \label{eqn:123898}
\frac{\gamma}{\Pr[\text{food remains for agent $i$}]},
\end{align}
so that the unconditional probability of agent $i$ being offered food is exactly
\begin{align*}
\Pr[\text{food remains for agent $i$}]\cdot \frac{\gamma}{\Pr[\text{food remains for agent $i$}]}=\gamma.
\end{align*}
Therefore, given any $\gamma$, the rationing policy is determined.  The optimal promise is then the maximum $\gamma$ for which this determined policy is feasible, which requires the probability in~\eqref{eqn:123898} to be at most 1.
This is established in the \namecref{thm:selection} below.

\begin{theorem} \label{thm:selection}
For any instance of \Cref{prob:selection} defined by $k=1$ and vector $(x_i)_{i\in[n]}$, \Cref{alg:selection} with $\gamma=1/(1+\sum_{i<n}x_i)$ is the optimal policy and guarantee.
\end{theorem}

\begin{algorithm}[H]
\caption{for \Cref{prob:selection} assuming $k=1$, parameterized by $\gamma>0$}\label{alg:selection}
\begin{algorithmic}
\For{agents $i=1,\ldots,n$} 
\State Draw an independent random bit $A_i$ that is 1 w.p.~$\gamma/(1-\gamma\sum_{j<i}x_{j})$
\If{food remains \textbf{and} $A_i=1$}
\State Offer food to agent $i$
\EndIf
\EndFor
\end{algorithmic}
\end{algorithm}

\proof{Proof of \Cref{thm:selection}.}
Let $\Offer_i,X_i\in\{0,1\}$ be the indicator random variables for agent $i$ being offered food and needing food, respectively.
In order to have a guarantee of $\gamma$, we must have $\bE[\Offer_i]=\gamma$ for all $i$.  However, $\Offer_n\le 1-\sum_{i<n}\Offer_i X_i$, because food can only be offered to agent $n$ if it was not already offered to an earlier agent who needed the food.  We get
\begin{align} \label{eqn:42810}
\gamma=\bE[\Offer_n]\le1-\sum_{i<n}\bE[\Offer_i]\bE[X_i]=1-\sum_{i<n}\gamma x_i,
\end{align}
applying both the linearity of expectation and the independence of $X_i$.
Rearranging~\eqref{eqn:42810} proves that the guarantee $\gamma$ cannot exceed $1/(1+\sum_{i<n}x_i)$.

We now prove that if \Cref{alg:selection} is run with parameter $\gamma=1/(1+\sum_{i<n}x_i)$, then $\bE[\Offer_i]=\gamma=1/(1+\sum_{i<n}x_i)$ for all $i$.  We induct over $i=1,\ldots,n$.  By definition, $\bE[\Offer_1]=\bE[A_1]=\gamma$.  For $i>1$, food is offered if and only if it still remains and $A_i=1$, so
\begin{align*}
\bE[\Offer_i]&=\bE\left[1-\sum_{j<i}\Offer_{j}X_{j}\right]\bE[A_i]
\\ &=(1-\sum_{j<i}\gamma x_{j})\frac{\gamma}{1-\gamma\sum_{j<i}x_{j}}
\end{align*}
where the first equality applies the independence of $A_i$; the second equality applies the independence of $X_{j}$ and the definition of $A_i$.  The final expression cancels to equal $\gamma$, completing the induction.

Finally, we must that verify \Cref{alg:selection} is feasible, by showing that the probability of random bit $A_i$ lies in [0,1] for all $i$.  Substituting in the value of $\gamma$, this probability can be re-written as
\begin{align*}
\bE[A_i]=\frac{1}{1/\gamma-\sum_{j<i}x_{j}}=\frac{1}{1+\sum_{i<n}x_i-\sum_{j<i}x_{j}}=\frac{1}{1+\sum_{j=i}^{n-1}x_{j}}
\end{align*}
which clearly satisfies $0\le\bE[A_1]\le\cdots\le\bE[A_n]=1$.
\Halmos\endproof

\subsection{What if $k>1$?} \label{sec:selectionK}

Food pantries should have starting supply $k$ greater than 1.  However, this significantly complicates the problem, illustrating perhaps the main intricacy of sequential randomized rounding.  Indeed, suppose $k=2$, and the mobile pantry is deciding whether to offer food to agent 2.  At this point, there would be either 1 or 2 units remaining, with
\begin{align*}
\Pr[\text{1 unit remaining}] &=\Pr[\text{agent~1 was offered food}]\Pr[\text{agent~1 needs food}]=\gamma x_1.
\end{align*}
To offer food to agent 2 w.p.~$\gamma$, the algorithm must define random bits $A_2',A_2''$ satisfying
\begin{align}
&& \Pr[\text{1 unit remaining}]\cdot\bE[A_2']+\Pr[\text{2 units remaining}]\cdot\bE[A_2'']=\gamma \nonumber
\\ \Longleftrightarrow && \gamma x_1\bE[A_2']+(1-\gamma x_1)\bE[A_2'']=\gamma \label{eqn:12839}
\end{align}
where $A_2',A_2''$ indicate whether to offer food to agent $2$ when there are $1,2$ units remaining, respectively.
The key difference from the $k=1$ case is even that for a fixed promise $\gamma$, the rationing policy is no longer determined---the algorithm has a choice on how to balance between the probabilities of $A_2'$ and $A_2''$.

\citet{alaei2011bayesian} also proposed a rule for handling this case---first try to satisfy~\eqref{eqn:12839} by fixing $\bE[A_2']=0$ and setting $\bE[A_2'']=\gamma/(1-\gamma x_1)$; if this is infeasible because $\gamma/(1-\gamma x_1)>1$, then instead set $\bE[A_2'']=1$ and increase $\bE[A_2']$ as needed to satisfy~\eqref{eqn:12839}.
This is saying that we would always rather offer food to agent 2 when there are 2 units remaining instead of 1 unit remaining; only if this is insufficient to promise agent 2 a probability $\gamma$ of being offered food  (because the probability of having both units remaining is too low), might we offer them the last remaining food supply.
Alaei's rule can be generalized to an arbitrary $k$, and is quite intuitive: be more willing offer food when there is more remaining, which is essentially the philosophy of rationing.

Alaei's rule is in fact optimal, as later proven by \citet{jiang2022tight}, for any instance defined by $k$, $n$, and $(x_i)_{i\in[n]}$.
They formalized Alaei's rule using the following LP:
\begin{subequations}\label{lp:kUnitOCRS}
\begin{align}
\max\quad &\gamma
\\ \text{s.t.}\quad &\gamma = \sum_{\ell=1}^k \alpha^\ell_i &\forall i\in[n]
\label{lp:kUnitOCRS:acceptance}
\\ &\beta^\ell_i =
\begin{cases}
1 & i=1,\ell=k \\
0 & i=1,\ell<k \\
\beta^\ell_{i-1}-(\alpha^\ell_{i-1}+\alpha^{\ell+1}_{i-1})x_{i-1} & i>1
\end{cases}
&\forall i\in[n],\ell\in[k] \label{lp:kUnitOCRS:update}
\\ &0\le \alpha^\ell_i \le \beta^\ell_i &\forall i\in[n],\ell\in[k]
% \label{lp:kUnitOCRS:active}
\end{align}
\end{subequations}
In this LP, $\alpha^\ell_i$ is a decision variable representing the probability to offer food to agent $i$ when there are $\ell$ units remaining.
Note that $\alpha^\ell_i$ is an unconditional probability; e.g.\ $\alpha^1_2$ would be correspond to the expression $\gamma x_1\bE[A_2']$ from~\eqref{eqn:12839} (trying to make $\alpha^1_2$ correspond to $\bE[A_2']$ would result in non-linearities).
Meanwhile, $\beta^\ell_i$ is a "state" variable representing the probability of having $\ell$ units remaining when driving by agent $i$, updated via the equations~\eqref{lp:kUnitOCRS:update}.  For an agent $i>1$, we need to remove mass from $\beta^\ell_i$ if we had $\ell$ units for agent $i-1$ and they took one, and add mass to $\beta^\ell_i$ if we had $\ell+1$ units for agent $i-1$ and they took one, reflecting the final equation in~\eqref{lp:kUnitOCRS:update} (in which $\alpha^{k+1}_{i-1}$ is understood to be 0).
Finally, $\gamma$ is a free variable tuned to be the optimal guarantee.

\begin{theorem}[\citet{jiang2022tight}] \label{thm:selectionK}
For any instance of \Cref{prob:selection} defined by $k$, $n$, and $(x_i)_{i\in[n]}$, the optimal guarantee $\gamma$ is found by solving LP~\eqref{lp:kUnitOCRS}.  Moreover, there exists an optimal solution corresponding to Alaei's rule, found by defining $(\beta^\ell_i)_{\ell\in[k]}$ via~\eqref{lp:kUnitOCRS:update} and then $(\alpha^\ell_i)_{\ell\in[k]}$ via
\begin{align*}
\alpha^k_i= \min\{\gamma,\beta^\ell_i\}, \quad \alpha^{k-1}_i= \min\{\gamma-\alpha^k_i,\beta^\ell_i\},\quad\ldots,\quad\alpha^1_i=\min\{\gamma-\sum_{\ell>1}\alpha^\ell_i,\beta^\ell_i\}
\end{align*}
over iterations $i=1,\ldots,n$.  The corresponding optimal policy is: for agents $i=1,\ldots,n$, conditional on the number of units remaining being $\ell\in[k]$, offer one to the agent w.p.~$\alpha^\ell_i/\beta^\ell_i$.
\end{theorem}

\begin{remark}[State Tracking] \label{rem:intricacies}
The food rationing problem with $k>1$ is a good way to highlight two general intricacies about sequential randomized rounding:
\begin{itemize}
\item The algorithm should track the probabilities of the parallel states that it could be in at any given time
(achieved in this problem via the $\beta^\ell_i$ variables in~\eqref{lp:kUnitOCRS:update}, where $\ell$ is state and $i$ is time);
\item The action to take in one state depends on the hypothetical action that would have been taken in another parallel state (achieved in this problem via~\eqref{lp:kUnitOCRS:acceptance}, which couples the $\alpha^\ell_i$ variables across $\ell$).
\end{itemize}
\end{remark}

In dynamic programming, updating Bellman's equations requires knowing the value-to-go from future states, but not the probability of being on the current sample path, let alone the probabilities and actions for parallel hypothetical sample paths that the algorithm could be in at the current time.
Therefore, the two intricacies in \Cref{rem:intricacies} may seem counter-intuitive from a dynamic programming perspective for online stochastic optimization.
However, they are crucial from the randomized rounding perspective, and in particular crucial for this problem where there is a probabilistic fairness constraint.

\subsection{What if the agents can be visited in a different order?} \label{sec:selectionRandomOrder}

Although the optimal rationing policy can now be found, the fairness constraint that every agent must be offered food with the same probability results in a loss in efficiency.  To illustrate this, we return to the earlier example where there was $k=1$ unit and $n=2$ agents with $x_1=x_2=1/2$.  We saw that the optimal fair policy has guarantee $\gamma=2/3$, which means that the food gets taken and utilized w.p.~$\gamma(x_1+x_2)=2/3$.  Alternatively, the policy that never skips the first agent utilizes the food w.p.~$x_1+(1-x_1)x_2=3/4$, which is higher and hence more efficient.

In this example, the mobile pantry can in fact get the best of both worlds, if on half of the days, if reversed its route and offered the food to agent 2 first.  Indeed, each agent would then be offered food w.p.~$\gamma=\frac{1+1/2}2=3/4$, which in this case means that the utilization probability is also $3/4(x_1+x_2)=3/4$.
Of course, in general it could be costly to re-route, so it is useful to quantify the gain from being able to visit agents in different orders on different days.

To quantify this, we define the efficiency to be $\gamma$, because subject to the fairness constraint that every agent is offered food with the same probability $\gamma$, the expected utilization is proportional to $\gamma$.
For an instance defined a vector $\vx=(x_i)_{i\in[n]}\in[0,1]^n$ of arbitrary length $n$, we let $\gamma_k(\vx)$ denote the maximum $\gamma$ possible when the mobile pantry produces $k$ units and visits the agents in order $i=1,\ldots,n$, noting that $\gamma_k(\vx)$ equals the optimal objective value of LP~\eqref{lp:kUnitOCRS}.  Meanwhile, we let $\gRO_k(\vx)$ denote the maximum $\gamma$ possible when the mobile pantry produces $k$ units and visits the agents in a uniformly random order.
We are interested in how $\gamma_k(\vx)$ compares to $\gRO_k(\vx)$; e.g.\ earlier we showed that $\gamma_1(1/2,1/2)=2/3$ while $\gRO_1(1/2,1/2)=3/4$.

Following the literature, we compare the values of $\gamma_k(\vx)$ and $\gRO_k(\vx)$ under worst-case instances.
Although not critical for the results, we restrict to instances where $\sum_{i=1}^nx_i\le k$, an assumption that the supply meets the expected demand.  (This will always be satisfied
when we apply these results beyond fairness problems in \Cref{sec:matching}.)
For a fixed $k$, we compare the measures $\gamma_k$ and $\gRO_k$, defined as
\begin{align*}
\gamma_k = \inf_n \inf_{\vx\in[0,1]^n:\sum_i x_i \le k} \gamma_k(\vx);
\\ \gRO_k = \inf_n \inf_{\vx\in[0,1]^n:\sum_i x_i \le k} \gRO_k(\vx).
\end{align*}
Although it is perhaps more useful compare $\gamma_k(\vx)$ to $\gRO_k(\vx)$ for the specific instance $\vx$ in practice, the theoretical canon is to compare worst-case measures like $\gamma_k$ and $\gRO_k$, because they are not specific to the application at hand and not sensitive to a change in the instance.  In this particular problem, it is also unknown how to compute $\gRO(\vx)$; there is no known LP formulation like~\eqref{lp:kUnitOCRS}.

We now derive the values of $\gamma_1$ and $\gRO_1$.  \Cref{thm:selection} says that $\gamma_1(\vx)=1/(1+\sum_{i<n}x_i)$ for any instance $\vx$, from which it is easy to see that $\gamma_1(\vx)\ge 1/2$ as long as $\sum_i x_i\le 1$, and that this lower bound is attained, implying $\gamma_1=1/2$.  For $\gRO_1$, although it is unknown how to compute $\gRO(\vx)$, it can be lower-bounded using an algorithm of \citet{lee2018optimal}.  They employ a well-known trick for analyzing random orders, which is to model every agent as arriving at a random "time" drawn independently and uniformly from [0,1].  This will later allow for a clever integration in the analysis.

\begin{remark}
If the agents arrive in a uniformly random order without arrival times, then they can be assigned arrival times using the following process.  Let Unif[0,1] denote the uniform distribution over [0,1].  The process draws random variables $U_1,\ldots,U_n$ IID from Unif[0,1] and places them in a stack, sorted so that the smallest number is on top and the largest number is on the bottom.  Whenever an agent $i$ arrives in the uniformly random order, we pretend that their arrival time is the (smallest) number on the top of the stack, after which we remove this number from the stack.  

To see that the agents end up with IID Unif[0,1] arrival times, let $\sigma:[n]\to[n]$ denote the random permutation such that the first agent to arrive has index $\sigma(1)$, ..., the last agent to arrive is $\sigma(n)$.
The distribution of arrival times is unchanged in an alternate world where $\sigma$ is drawn first, and then the random variables $U_1,\ldots,U_n$ are drawn conditional on $U_{\sigma(1)}<\cdots<U_{\sigma(n)}$.
In this alternate world, agent $\sigma(1)$ arrives at time $U_{\sigma(1)}$, ..., agent $\sigma(n)$ arrives at time $U_{\sigma(n)}$.
That is, irrespective of $\sigma$
each agent $i=1,\ldots,n$ always gets assigned the arrival time $U_i$, which is an independent Unif[0,1] random variable.
Because this alternate world has the same distribution of arrival times as the original process, this shows that each agent $i$ indeed gets assned an independent Unif[0,1] arrival time.
\end{remark}

\begin{theorem}[\citet{lee2018optimal}] \label{thm:selection2}
For any instance of \Cref{prob:selection} satisfying $\sum_{i=1}^n x_i\le k=1$, \Cref{alg:selection2} offers food to every agent w.p.\ at least\footnote{We are technically changing the problem formulation here, allowing agents to be offered food w.p.\ greater than $\gamma$.  This is actually the standard problem formulation, which is equivalent in theory because an optimal policy never wants to offer food to an agent w.p.\ greater than the required $\gamma$.} $1-1/e$, assuming agents are encountered in a uniformly random order.
\end{theorem}

\begin{algorithm}[H]
\caption{for \Cref{prob:selection} assuming $\sum_{i=1}^n x_i\le k=1$ and random order}\label{alg:selection2}
\begin{algorithmic}
\State Draw time $U_i$ independently for each agent $i$, uniformly from [0,1]
\For{agents $i=1,\ldots,n$ in increasing order of $U_i$} 
\State Draw an independent random bit $A_i$ that is 1 w.p.~$e^{-U_i x_i}$
\If{food remains \textbf{and} $A_i=1$}
\State Offer food to agent $i$
\EndIf
\EndFor
\end{algorithmic}
\end{algorithm}

\proof{Proof of \Cref{thm:selection2}.}
Follow the same notation as in the proof of \Cref{thm:selection}.  For any agent $i$,
\begin{align*}
\bE[\Offer_i]
&=\int_0^1 \bE[\Offer_i|U_i=t] dt
\\ &=\int_0^1 \bE\Big[A_i\prod_{j\neq i}(1-\bI(U_j<t)A_jX_j)\Big| U_i=t\Big] dt
\\ &=\int_0^1 \bE[A_i|U_i=t] \prod_{j\neq i}(1-\int_0^{t} \bE[A_j|U_j=u]\bE[X_j] du) dt
\\ &=\int_0^1 e^{-tx_i} \prod_{j\neq i}(1-\int_0^{t} e^{-ux_j}x_j du) dt
\\ &=\int_0^1 e^{-tx_i} \prod_{j\neq i}(1-(1-e^{-tx_j})) dt
\\ &=\int_0^1 e^{-t\sum_{i=1}^n x_i} dt
\\ &\le\int_0^1 e^{-t}dt.
% =1-1/e.
\end{align*}
The second equality holds because when agent $i$ is encountered, food remains if and only if no other agent $j$ who has arrived beforehand has $A_jX_j=1$, in which case agent $i$ is offered the food whenever $A_i=1$.
The third equality holds by independence.
The remaining equalities hold by definitions of random variables and elementary operations.
The inequality holds because $\sum_{i=1}^n x_i\le 1$, and the final integral can be evaluated to equal $1-1/e$, completing the proof.
\Halmos\endproof

\Cref{thm:selection2} shows that when $k=1$, assuming supply meets expected demand, every agent can be offered food w.p.\ at least $1-1/e\approx63.2\%$ if we visit them in a random order.  By contrast, if we visit them in a fixed order, then this guarantee might be as low as 50\%.  Hence, the benefit of switching to random order is at least $63.2\%-50\%=13.2\%$, which can then be compared to the logistical cost of re-routing the mobile pantry on different days.  Of course, the number 13.2\% serves as a rough benchmark, and is subject to the aforementioned caveat about worst-case analysis.  Regardless, \Cref{thm:selection} showed that any instance in which $\sum_{i<n} x_i$ is close to 1 will have $\gamma_1(\vx)$ not much better than 50\%, and \Cref{thm:selection2} shows that $\gRO(\vx)$ is at least 63.2\%, so this combined can be used to make an argument in favor of switching to random order.

When $k>1$, there is no closed-form for $\gamma_k$, but it is characterized in \citet{jiang2022tight}.
For example, $\gamma_2$ is the larger real number satisfying $e^{1/\gamma_2-3}=2/\gamma_2-3$, which is approximately $61.5\%$.
Regardless, one difference from $k=1$ is that worst-case instances now require arrivals to follow a Poisson process, which means that they may be further away from a typical instance.
Thus, one can alternatively just use LP~\eqref{lp:kUnitOCRS} to compute $\gamma_k(\vx)$ for the instance $\vx$ at hand, and a lower-bound on $\gRO_k(\vx)$ is then given by the following \namecref{thm:selection2K}.

\begin{theorem}[\citet{arnosti2022tight,lee2018optimal}] \label{thm:selection2K}
For any instance of \Cref{prob:selection} satisfying $\sum_{i=1}^n x_i\le k$, there exists a policy that offers food to every agent w.p.~$1-e^{-k}k^k/k!$, assuming agents are encountered in a uniformly random order.
\end{theorem}

$1-e^{-k}k^k/k!$ is actually the correct value of $\gRO_k$, as we later explain in \Cref{sec:explainingGuarantee}.  In particular, $\gRO_1=1-1/e$ (i.e.\ \Cref{alg:selection2} generates an optimal policy for a worst-case instance), and $\gRO_2=1-2e^{-2}\approx 72.9\%$.
However, there is no explicit policy behind \Cref{thm:selection2K}---it follows from applying an equivalence result of \citet{lee2018optimal} on an "ex-ante prophet inequality" of \citet{arnosti2022tight}.  This illustrates just one of the many subtleties that make the class of problems studied in this \namecref{sec:selection} interesting.
% We wrap up by providing further references.

\subsection{Further References on OCRS} \label{sec:ocrs}

The algorithms studied in this \namecref{sec:selection} are called Online Contention Resolution Schemes (OCRSs) in the literature.
In this literature, agents are said to be "arriving", instead of being visited by a mobile food pantry.
A more salient difference is that when an agent $i$ arrives, they first declare whether they need food (occurring independently w.p.~$x_i$), in which case their state is said to be "active".  Only for active agents is there an option to give food, in which case they are said to be "accepted", and the objective is to accept every agent w.p.~$\gamma x_i$.  Regardless, our \Cref{prob:selection} is equivalent if we imagine the offering of food to an agent as deciding, before seeing their state, to accept them whenever they are active.
For every agent $i$, being offered food w.p.~$\gamma$ is then equivalent to being accepted w.p.~$\gamma x_i$ (by independence).

In this \namecref{sec:selection} we had focused on the constraint where at most $k$ agents can be accepted, but we now describe the OCRS problem for a general feasibility constraint.  The subset of accepted agents is constrained to lie in $\cF$, a downward-closed collection of subsets of $[n]$.  Vector $\vx$ is restricted to lie in a polytope $P_\cF\subseteq[0,1]^n$ that contains the indicator vector $(\bI(i\in S))_{i\in[n]}$ of any feasible $S\in\cF$.  The agents $i$ arrive in some order, each active independently w.p.~$x_i$, and the goal is to accept every agent w.p.\ at least $\gamma x_i$. The maximum such $\gamma$ that can be achieved for all $\cF$ in some class and any $x\in P_\cF$, under a particular arrival order, is referred to as the selectability (for that class and arrival order).

Cast in this language, we had studied the feasibility class where $\cF=\{S\subseteq[n]:|S|\le k\}$ and $P_\cF=\{\vx\in[0,1]^n:\sum_i x_i\le k\}$, for some given $k$ and arbitrary $n$.
We showed that the selectability is $\gamma_k$ under fixed arrival order $i=1,\ldots,n$, and $\gRO_k$ under random arrival order.
The feasibility class that we had studied is sometimes called a $k$-uniform matroid, and we now mention three other classes.
First, pioneering work on OCRS (\citet{feldman2021online}) focused on the feasibility class of general matroids, for which the selectability is known to be $1/2$ under fixed order (\citet{kleinberg2012matroid,lee2018optimal}) and $1-1/e$ under random order (\citet{ehsani2018prophet,lee2018optimal}).
We defer the definition of general matroids and the matroid polytope to those papers.
Second, in the knapsack feasibility class, each agent $i$ has a size $d_i\in[0,1]$ and a subset of agents is feasible if and only if their total size does not exceed 1, i.e.~$\cF=\{S\subseteq[n]:\sum_{i\in S}d_i\le1\}$.  When $P_\cF$ is defined to be the knapsack polytope $\{\vx\in[0,1]^n:\sum_i d_ix_i\le 1\}$, the selectability under fixed order is known to be $1/(3+e^{-2})\approx0.319$ (\citet{jiang2022tight}).  No results are known for random order.
Third, a feasibility class of interest is matchings in graphs, which is an active area of research for OCRSs in which few tight results are known.  We defer defining this problem to \Cref{sec:ocrsGraphs}, because it is related to stochastic matching in graphs.

We finish with references to a few other variants and applications of OCRS that are beyond the scope of this tutorial.
First, contention resolution originated in an offline setting, motivated by randomized rounding for submodular optimization (\citet{chekuri2014submodular}).
There, an important property for algorithms to satisfy is monotonicity, which we do not consider.
We also do not consider the greedy property for OCRS, which is needed if one wants to play against an arrival order chosen by the strongest, almighty adversary, as discussed in \citet{feldman2021online}.
The stated results for matching (\citet{ezra2022prophet}) and $k$-uniform matroid/knapsack (\citet{jiang2022tight}) hold against the oblivious and online adversary respectively, which are also discussed in \citet{feldman2021online}.
Finally, it is interesting to consider contention resolution with correlated activeness probabilities, as investigated in \citet{dughmi2019outer}.

\section{Hiring the Best Candidate} \label{sec:hiring}

In \Cref{sec:selection} we studied online contention resolution, which was a sequential randomized rounding problem by definition.  In this \namecref{sec:hiring} we study a stochastic dynamic programming problem from \citet{purohit2019hiring} that admits a simple randomized rounding algorithm.

\begin{problem}[Sequential Offering] \label{prob:hiring}
A firm has $n$ candidates whom it would like to hire for one of $k<n$ identical positions.
It has determined for each candidate $i$ a value ("weight") $w_i\ge0$ and their probability $p_i$ of accepting an offer.
The firm has time to make $T$ offers sequentially, where it is assumed that each candidate immediately decides whether to accept the offer, independent of other candidates.  The objective is to maximize performance, which is defined as the expected total weight of candidates hired.
\end{problem}

To build some intuition, note that if $T$ is as large as $n$, then the optimal policy is to make offers in decreasing order of $w_i$.
Indeed, because the firm will never run out of time, it should go for the highest-valued candidates first, irrespective of their probability of accepting.
On the other hand, if $T$ is as small as $k$, then the optimal policy is to make offers in decreasing order of $w_i p_i$.
Indeed, because the firm can make only one offer for each position, it should maximize the expected return from each offer, which is $w_i p_i$.
In the interesting case where $k<T<n$, the optimal sequential hiring policy is non-trivial, and a naive dynamic program would also fail because the state space, described by the set of candidates yet to receive an offer, is exponential in $n$.

Nonetheless, we can derive a simple algorithm for this problem via randomized rounding.  We write the following LP to guide our decisions:
\begin{subequations} \label{lp:hiring}
\begin{align}
\max\quad & \sum_{i=1}^n w_i p_i y_i 
\\ \mathrm{s.t.}\quad & \sum_{i=1}^n y_i \le T \label{lp:hiring:T}
\\ & \sum_{i=1}^n p_i y_i \le k \label{lp:hiring:k}
\\ & 0\le y_i\le 1 &\forall i\in[n]
\end{align}
\end{subequations}
In this LP, $y_i$ is a decision variable representing the probability of making an offer to candidate $i$.  Constraints~\eqref{lp:hiring:T} and~\eqref{lp:hiring:k} impose the expected number of offers to not exceed $T$, and the excepted number of accepted offers to not exceed $k$.

\begin{lemma} \label{lem:hiring}
For any instance of \Cref{prob:hiring}, the performance of any policy is at most $\LP$.
\end{lemma}

\proof{Proof of \Cref{lem:hiring}.}
Fix any instance and policy for \Cref{prob:hiring}. The performance of the policy equals the expected total weight of the candidates it hires.
Let $Y_i\in\{0,1\}$ indicate whether candidate $i$ is made an offer, and $P_i\in\{0,1\}$ indicate whether $i$ would accept an offer if made.
On any sample path, the policy's execution satisfies $\sum_i Y_i\le T$ and $\sum_i P_i Y_i \le k$, and its total weight hired equals $\sum_i w_i P_i Y_i$.
Taking expectations, we get $\sum_i \bE[Y_i]\le T$ and $\sum_i \bE[P_iY_i]\le k$, and that the policy's performance equals $\sum_i w_i \bE[P_i Y_i]$.
Crucially, $\bE[P_iY_i]=\bE[P_i]\bE[Y_i]=p_i\bE[Y_i]$ because $P_i$ is independent from $Y_i$, due to offers being made without knowing whether the candidate would accept (this argument requires the independence of acceptance decisions across candidates).
Thus, setting $y_i=\bE[Y_i]$ for all $i\in[n]$ defines a feasible solution to LP~\eqref{lp:hiring}, whose objective value equals the performance of the policy under consideration.  Therefore, the optimal objective value of the LP can only be higher, completing the proof.
\Halmos\endproof

We now show how to convert any feasible solution $(y_i)_{i\in[n]}$ for LP~\eqref{lp:hiring} into a policy for \Cref{prob:hiring}.
Whereas $(y_i)_{i\in[n]}$ only satisfied the constraints of $T$ offers and $k$ positions in expectation, the policy must satisfy them always.
A simple observation made in \citet{purohit2019hiring} is that given any fixed set of candidates to receive offers, it is optimal to send to those candidates $i$ in decreasing order of $w_i$, so that the highest-valued candidates have a chance to fill the positions first.
We will apply this observation after randomly rounding $(y_i)_{i\in [n]}$ into an offer set.
The algorithm we present is from \citet{epstein2024selection}, who observe that the rounding is actually very easy, given the simple structure of the LP.

\begin{theorem} \label{thm:hiring}
For any instance of \Cref{prob:hiring}, the performance of \Cref{alg:hiring} is at least $(1-e^{-k}k^k/k!)\LP$.
In particular, when $k=1$, this guarantee relative to $\LP$ is $1-1/e$.
\end{theorem}

\begin{algorithm}[H]
\caption{for \Cref{prob:hiring}}\label{alg:hiring}
\begin{algorithmic}
\State Solve LP~\eqref{lp:hiring}, letting $(y_i)_{i\in[n]}$ denote a basic optimal solution
\State Set $Y_i=y_i$ for all candidates $i$ with $y_i\in\{0,1\}$
\State Let $S=\{i:y_i\in(0,1)\}$ \Comment{in a basic solution, $|S|\le 2$}
\If{$|S|=1$}
\State Set $Y_i=1$ w.p.~$y_i$ and $Y_i=0$ otherwise, for the candidate $i\in S$
\ElsIf{$|S|=2$} \Comment{if $|S|=2$, then constraint~\eqref{lp:hiring:T} must be binding and hence $\sum_{i\in S} y_i=1$}
\State Set $Y_i=1$ for exactly one of the candidates $i\in S$, satisfying $\bE[Y_i]=y_i$ for both $i\in S$
\EndIf
\State Make offers to candidates $i$ with $Y_i=1$ in decreasing order of $w_i$, while positions remain
\end{algorithmic}
\end{algorithm}

\proof{Proof of \Cref{thm:hiring} when $k=1$.}
We first argue that \Cref{alg:hiring} is feasible, in that it never makes more than $T$ offers.  This is immediate if $|S|=0$.  If $|S|=1$, then $\sum_{i\notin S}y_i\le T-1$ by LP constraint~\eqref{lp:hiring:T}, which means that even if $Y_i$ is rounded up to 1 for the $i\in S$, we would have $\sum_{i=1}^n Y_i\le T$.  If $|S|=2$, then we know $\sum_{i\notin S}y_i=T-1$, and hence $\sum_{i=1}^n Y_i= T$ because exactly one of the candidates in $i\in S$ will have $Y_i$ rounded up to 1.

Now, re-index the candidates so that $w_1\ge\cdots\ge w_n$ and define $w_{n+1}=0$.  Letting $(y_i)_{i\in[n]}$ refer to the optimal solution from \Cref{alg:hiring}, we can express
\begin{align} \label{eqn:90178}
\LP=\sum_{i=1}^n w_i p_i y_i
=\sum_{i=1}^n \left(\sum_{j=i}^n(w_{j}-w_{j+1})\right) p_i y_i
=\sum_{j=1}^n(w_{j}-w_{j+1}) \sum_{i=1}^{j} p_i y_i
\end{align}
where each of the differences $w_j-w_{j+1}$ is non-negative.

Meanwhile, let $P_i\in\{0,1\}$ indicate whether candidate $i$ would accept an offer.  The policy hires a candidate $i\in[n]$ if and only if $P_iY_i=1$ (they are made an offer and accept), and $P_jY_j=0$ for all $j<i$ (none of the higher-valued candidates $j$ who came before $i$ were made an offer and accepted).  Following a manipulation analogous to~\eqref{eqn:90178}, we can express the policy's total weight as
\begin{align}
\sum_{i=1}^n w_i P_iY_i(1-P_{i-1}Y_{i-1})\cdots(1-P_1Y_1)
&=\sum_{j=1}^n (w_{j}-w_{j+1}) \sum_{i=1}^{j}P_iY_i (1-P_{i-1}Y_{i-1})\cdots(1-P_1Y_1) \nonumber
\\ &=\sum_{j=1}^n (w_{j}-w_{j+1}) \left(1-\prod_{i=1}^{j}(1-P_iY_i)\right) \label{eqn:78708}
\end{align}
which has an intuitive explanation: the difference $w_j-w_{j+1}$ is accumulated by the policy if and only if at least one of the candidates $i=1,\ldots,j$ (who have $w_i\ge w_j$) has $P_iY_i=1$.
We claim that these candidates having $P_iY_i=1$ is negatively correlated across $i$, which means that the probability of $w_j-w_{j+1}$ not being accumulated should be lower in than the independence case, or more formally:
\begin{align} \label{eqn:negCor}
\bE\left[\prod_{i=1}^{j}(1-P_iY_i)\right] \le \prod_{i=1}^{j}\bE[1-P_iY_i].
\end{align}
This can be argued from first principles: if $|S\cap[j]|<2$ then \eqref{eqn:negCor} is satisfied as equality due to independence; otherwise if $S=\{i',i''\}$ with $i',i''\le j$ then~\eqref{eqn:negCor} is equivalent to
$$
(y_{i'}(1-p_{i'})+y_{i''}(1-p_{i''}))\prod_{i\le j,i \neq i',i''}(1-p_iy_i)
\le (1-p_{i'}y_{i'})(1-p_{i''}y_{i''})\prod_{i\le j,i\neq i',i''}(1-p_iy_i)
$$
which is true due to $y_{i'}+y_{i''}=1$.
Given~\eqref{eqn:negCor}, the expectation of~\eqref{eqn:78708} is at least
\begin{align*}
\sum_{j=1}^n (w_{j}-w_{j+1}) \left(1-\prod_{i=1}^{j}(1-p_iy_i)\right)
&\ge\sum_{j=1}^n (w_{j}-w_{j+1}) \left(1-\prod_{i=1}^{j}e^{-p_iy_i}\right)
\\ &=\sum_{j=1}^n (w_{j}-w_{j+1}) \left(1-e^{-\sum_{i=1}^{j}p_iy_i}\right).
\end{align*}

It remains to compare this expression to $\LP$ as expressed in~\eqref{eqn:90178}.  By LP constraint~\eqref{lp:hiring:k}, we know that $\sum_{i=1}^j p_iy_i\le k=1$.
Because the function $(1-e^{-x})/x$ is decreasing in $x$, we can prove
$$\frac{1-e^{-\sum_{i=1}^{j}p_iy_i}}{\sum_{i=1}^j p_i y_i}\ge \frac{1-e^{-1}}{1}=1-1/e$$
for any $j\in[n]$ by letting $x=\sum_{i=1}^j p_i y_i$.  Finally, each of the differences $w_j-w_{j+1}$ is non-negative, so the policy's performance relative to $\LP$ is at least $1-1/e$, completing the proof when $k=1$.
\Halmos\endproof

We presented an elementary proof for \Cref{thm:hiring} that only works when $k=1$.
The result for general $k$ is proven in \citet{epstein2024selection} using the powerful tool of correlation gap (\citet{yan2011mechanism}).

\subsubsection{Explaining the $1-e^{-k}k^k/k!$ guarantee.} \label{sec:explainingGuarantee}

It is no coincidence that the guarantee for general $k$ is identical to what was found in \Cref{thm:selection2K}.  All of these results compare a relaxation that hires $k$ candidates in expectation to a reality that is subject to the randomness in whether each candidates accepts.  That is, letting $x_i=y_ip_i$ denote the probability of hiring candidate $i$, reality can only hire $\bE[\min\{\sum_{i=1}^n\mathrm{Ber}(x_i),k\}]$ candidates, where $\mathrm{Ber}(x_i)$ denotes an independent Bernoulli random variable with mean $x_i$.  
These results eventually argue that the worst case occurs when there are an infinite number of candidates with infinitesimal probabilities:
\begin{align}
\inf_n \inf_{\vx\in[0,1]^n:\sum_i x_i\le k}\frac{\bE[\min\{\sum_{i=1}^n\mathrm{Ber}(x_i),k\}]}{\sum_i x_i}
&=\lim_{n\to\infty}\frac{\bE[\min\{\mathrm{Bin}(n,\frac kn),k\}]}{k} \label{eqn:corrGap}
\\ &=\frac{\bE[\min\{\mathrm{Pois}(k),k\}]}{k} \label{eqn:corrGap2}
\\ &=1-e^{-k}\frac{k^k}{k!} \label{eqn:corrGap3}
\end{align}
where $\mathrm{Bin}(n,\frac kn)$ denotes a Binomial random variable with $n$ trials of probability $\frac kn$
and
$\mathrm{Pois}(k)$ denotes a Poisson random variable with mean $k$.
A proof of the very useful sequence of facts in~\eqref{eqn:corrGap}--\eqref{eqn:corrGap3} (assuming $\sum_{i=1}^n x_i=k$) can be found in \citet[Lemma~4.2]{yan2011mechanism}.

\Cref{eqn:corrGap,eqn:corrGap2,eqn:corrGap3} also suggest that the guarantee of $1-e^{-k}k^k/k!$ is best possible, or "tight", relative to the value of $\LP$.
Indeed, one can construct the following sequence of instances.

\begin{example}
Let $T=n$, and $w_i=1,p_i=k/n$ for all candidates $i\in[n]$.
Setting $y_i=1$ for all $i$ is feasible and hence optimal for LP~\eqref{lp:hiring}, with $\LP=k$.
Meanwhile, any policy would make offers to the symmetric candidates in an arbitrary order until either all $n$ candidates have been offered or all $k$ positions have been filled, whichever happens first.
The performance of the policy is $\bE[\min\{\mathrm{Bin}(n,k/n),k\}]$.
Taking $n\to\infty$, the policy's performance divided by $k$ is $1-e^{-k}k^k/k!$ due to~\eqref{eqn:corrGap}--\eqref{eqn:corrGap3}, and hence it is not possible to have a guarantee better than $1-e^{-k}k^k/k!$ relative to $\LP$.
\end{example}

This also applies to the fairness problem from \Cref{sec:selection}, if we argue that promising every agent probability $\gamma$ of being offered food would result in $\gamma\sum_{i=1}^n x_i$ food being utilized in expectation, which cannot exceed $\bE[\min\{\sum_{i=1}^n\mathrm{Ber}(x_i),k\}]$.  \Cref{eqn:corrGap,eqn:corrGap2,eqn:corrGap3} then show that $\gamma$ must be as small as $1-e^{-k}k^k/k!$ on a worst-case instance satisfying $\sum_{i=1}^n x_i\le k$.

Finally, we note that
\begin{align*}
1-e^{-k}\frac{k^k}{k!}\approx 1-\frac1{\sqrt{2\pi k}}
\end{align*}
by Stirling's approximation of $k!$, which means that the guarantee is approaching 100\% as $k\to\infty$.  This is intuitive: because each candidate accepts following an independent random variable, the advantage of the relaxation over reality vanishes as $k$ grows, by the law of large numbers.

\subsubsection{Discussion about negative correlation.}
\Cref{sec:explainingGuarantee} suggests that the constraint of $T$ offers was inconsequential for the guarantee relative to the LP,
and all of the loss was coming from the constraint of $k$ positions.
This is indeed the case---\Cref{alg:hiring} "losslessly" rounds the solution $(y_i)_{i\in[n]}$ for LP~\eqref{lp:hiring} into variables $(Y_i)_{i\in[n]}\in\{0,1\}^n$ that satisfy $\sum_{i=1}^n Y_i\le T$.
Lossless here means two things:
first, the marginals are preserved in that $\bE[Y_i]=y_i$ for all $i$;
second, the $(Y_i)_{i\in[n]}$ variables are negatively correlated in the simplest and strongest form possible, where at most two of them are random, and if exactly two are random, then exactly of them will realize to 1.
Negative correlation allows us to argue (in~\eqref{eqn:negCor}) that the algorithm's performance in reality is no worse than that in an alternate reality where agents are sent offers independently at random with the same probabilities.

A weaker form of negative correlation that often arises in randomized rounding---see (P3) in \citet{gandhi2006dependent}---is sufficient for establishing our claim in~\eqref{eqn:negCor} in the proof for $k=1$.
However, the stronger form appears necessary for $k>1$, and we refer to \citet{qiu2022submodular} for a recent reference on how these different forms of negative correlation forms relate.

\subsubsection{Discussion about adaptivity gap.}
After all this, we should reveal that \Cref{prob:hiring} actually has a polynomial-time dynamic programming algorithm when $k=1$.
In fact, for any $k$, this dynamic program efficiently computes the optimal \textit{non-adaptive} policy, which fixed the order for making offers beforehand.
\citet{purohit2019hiring} show that there is no loss from fixing the order when $k=1$, and hence the dynamic program is optimal.

On the other hand, there is a loss when $k>1$.  Indeed, conditional on there being positions remaining after a set of offers, the exact number remaining is now random, depending on how many of those offers were accepted.
The number of positions remaining determines how the firm should prioritize between sending "longshot" offers, to candidates $i$ with high $w_i$ but low $p_i$, vs.\ "safe" offers, to candidates $i$ with high expected value $w_i p_i$.
Since this number cannot be determined beforehand, the dynamic program that fixes the ordering is not optimal, and \Cref{thm:hiring} provides the state-of-the-art approximation ratio when $k>1$.

Importantly, \Cref{alg:hiring} actually generates a non-adaptive policy, whose order is fixed beforehand based on the $(Y_i)_{i\in [n]}$ variables.
\Cref{thm:hiring} shows that its performance is at least $(1-e^{-k}k^k/k!)\LP$, and hence a non-adaptive\footnote{Technically our non-adaptive policy is randomized, but it can easily be de-randomized by trying both possibilities.} policy can have performance at least $1-e^{-k}k^k/k!$ times that of the best adaptive policy.
This comparison is referred to as the \textit{adaptivity gap} in the literature (e.g.~\citet{dean2008approximating,purohit2019hiring}), and is sensible in any problem where the order can be freely decided.  We study such problems in Sections~\ref{sec:hiring}, \ref{sec:stocKnap}, and \ref{sec:stocProbMatching} of this tutorial.
% , which includes the hiring problems studied in this \namecref{sec:hiring}, the stochastic knapsack problems studied in \Cref{sec:stocKnap}, and the search problems discussed in \Cref{sec:stocProbMatching}.

\subsection{Extensions: Constrained Free-order Prophets and ProbeTop-$k$} \label{sec:ptk}

We extend \Cref{thm:hiring} to a more general hiring problem where weights are real-valued, and applicants must be interviewed to discover their true weight.

\begin{problem}[Constrained Free-order Prophets, ProbeTop-$k$] \label{prob:hiring2}
A firm has $n$ applicants for $k<n$ identical positions.
Each applicant $i$ has a weight $W_i\ge0$ that can be determined upon an interview.
The prior distribution of $W_i$ is known (e.g.\ based on the applicant's resume) and independent across $i$.
The firm has time to conduct $T$ interviews, and applicants can only be hired after an interview.
In this model, the firm can directly hire applicants because they are assumed to always\footnote{This can capture adaptive policies for \Cref{prob:hiring} if for all $i$, we set $W_i$ equal $w_i$ w.p.~$p_i$ and 0 otherwise.  Indeed, applicant $i$ would always be rejected if $W_i=0$, and an optimal adaptive policy would not be interviewing $i$ next if $i$ would not be accepted conditional on $W_i=w_i$ (\citet{epstein2024selection}).  Therefore, this is equivalent to the model where applicants $i$ have deterministic weights $w_i$ and decide whether to accept offers w.p.~$p_i$.} accept an offer.
The objective is to maximize performance, which is defined as the expected total value of applicants hired.
\end{problem}

One might realize that there are two variants to \Cref{prob:hiring2}---does the firm need to to immediately make a hiring decision after each interview, or can it finish all $T$ interviews and then hire the $k$ applicants with the highest realized values?
The former variant is called the Constrained\footnote{This variant is interesting even without the constraint of $T$ interviews, in which case it is called the Free-order Prophet problem.} Free-order Prophet problem, while the latter variant is called the ProbeTop-$k$ problem.
Clearly the firm is better off in the ProbeTop-$k$ variant, but deriving optimal policies under either variant is interesting.
In fact, there are further variants depending on whether the interview order can be decided adaptively.

Randomized rounding derives policies for all of these variants in one fell swoop.
It generates a hiring policy that satisfies the strictest variant (interview order decided in advance, hiring decision made immediately after each interview), yet is comparable to policies under the most lenient variant (interview order decide adaptively, hiring decisions made at end).
Indeed,
letting $F_i$ denote the CDF of weight $W_i$ for all $i$,
we can write the following mathematical program:
\begin{subequations} \label{lp:hiring2}
\begin{align}
\max\ & \sum_{i=1}^n z_i\int_{1-x_i/z_i}^1 F^{-1}_i(q)dq \label{lp:hiring2:obj}
\\ \mathrm{s.t.\ } & \sum_{i=1}^n z_i \le T
\\ & \sum_{i=1}^n x_i \le k
\\ & 0\le x_i\le z_i\le 1 &\forall i\in[n]
\end{align}
\end{subequations}
Here, $z_i$ and $x_i$ are decision variables representing the probabilities of interviewing and hiring applicant $i$, respectively.
Thus, $x_i/z_i$ represents the probability of hiring applicant $i$ conditional on interviewing them.
For any $i$, if the probability of hiring applicant $i$ conditional on interviewing them is $p$, then their expected weight conditional on being hired cannot exceed
\begin{align} \label{eqn:quantile}
\frac1{p}\int_{1-p}^1 F^{-1}_i(q)dq,
\end{align}
for any $p\in[0,1]$.  To explain~\eqref{eqn:quantile}, note that the best the policy can do is hire applicant $i$ whenever their weight, after being interviewed, realizes to an outcome in the top $p$ fraction of scenarios.  But \eqref{eqn:quantile} denotes precisely the average value of $W_i$ over these scenarios, where $F^{-1}_i(\cdot)$ denotes the inverse CDF of the distribution of $W_i$, and $q$ denotes the "quantile" with $F^{-1}_i(q)$ being increasing in $q$.  We defer a formal definition of $F^{-1}_i(\cdot)$ to \citet{chawla2014bayesian}, and instead illustrate it through two examples: 
if $W_i$ is uniformly distributed over [0,1], then the CDF of $W_i$ is $F_i(w)=w$ and hence $F^{-1}_i(q)=q$ for all $q\in[0,1]$, which means~\eqref{eqn:quantile} equals $\frac{1+(1-p)}{2}=1-p/2$;
if $W_i$ is a discrete distribution equally likely to be 1 or 1/2, then~\eqref{eqn:quantile} equals $1$ for $p\le1/2$, and equals $\frac{(1/2)1+(p-1/2)1/2}{p}=1/2+1/(4p)$ for $p\ge1/2$.

We design an interviewing policy for \Cref{prob:hiring2} by reducing to \Cref{prob:hiring}.
To elaborate, fix an optimal solution $(x_i,z_i)_{i\in[n]}$ to~\eqref{lp:hiring2}, and we would like to compare against the value of objective~\eqref{lp:hiring2:obj}.
We force the interviewing policy to immediately hire an applicant $i$ if and only if their weight, upon being interviewed, realizes to a value in its top-$(x_i/z_i)$ quantile.
Interviewing an applicant $i$ is now equivalent to making an offer to candidate $i$ in an instance of \Cref{prob:hiring} with the same $k,T,n$ and
\begin{align*}
p_i=\frac{x_i}{z_i}, \qquad w_i=\frac1{p_i}\int_{1-p_i}^1 F^{-1}_i(q)dq && \forall i\in[n].
\end{align*}
Indeed, interviewing and making an offer would both result in a hire independently w.p.~$p_i$, conditional on which the weight $W_i$ is a random draw over the top $p_i$ fraction of scenarios, with expectation $w_i$.
Therefore, the sequential offering \Cref{alg:hiring} implies a sequential interviewing policy for \Cref{prob:hiring2} with performance at least $(1-e^{-k} k^k/k!)$ times the optimal objective value of LP~\eqref{lp:hiring} on the constructed instance of \Cref{prob:hiring}.
And because setting $y_i=z_i \forall i\in[n]$ forms a feasible solution, the optimal objective value of this LP~\eqref{lp:hiring} must be at least $\sum_{i=1}^n w_i p_i z_i$, which equals~\eqref{lp:hiring2:obj} by construction, completing the argument that the interviewing policy has performance at least $1-e^{-k}k^k/k!$ times the optimal objective value of~\eqref{lp:hiring2}.
The latter is an upper bound on the performance of any policy under the most lenient variant of interviewing, implying a strong result for adaptivity gaps.
% We design an interviewing policy for \Cref{prob:hiring2} by reducing to \Cref{prob:hiring}.
% To elaborate, fix an optimal solution $(x_i,z_i)_{i\in[n]}$ to~\eqref{lp:hiring2}, and we would like to compare against the value of objective~\eqref{lp:hiring2:obj}.
% We consider an interviewing policy that immediately hires an applicant $i$ if and only if their weight, upon being interviewed, realizes to a value in its top-$(x_i/z_i)$ quantile.
% Interviewing an applicant $i$ is now equivalent to making an offer to candidate $i$ in an instance of \Cref{prob:hiring} with the same $k,T,n$ and
% \begin{align*}
% p_i=\frac{x_i}{z_i}, \qquad w_i=\frac1{p_i}\int_{1-p_i}^1 F^{-1}_i(q)dq && \forall i\in[n].
% \end{align*}
% Indeed, interviewing and making an offer would both result in a hire independently w.p.~$p_i$, conditional on which the weight $W_i$ is a random draw over the top $p_i$ fraction of scenarios, with expectation $w_i$.
% Therefore, we can run the sequential offering \Cref{alg:hiring} with $(y_i)_{i\in[n]}=(z_i)_{i\in[n]}$ (a feasible solution to~\eqref{lp:hiring}) to get performance at least
% \begin{align*}
% (1-e^{-k} \frac{k^k}{k!})\sum_{i=1}^n w_ip_iy_i=(1-e^{-k} \frac{k^k}{k!})\sum_{i=1}^n z_i \int_{1-x_i/z_i}^1 F^{-1}_i(q)dq 
% \end{align*}
% by \Cref{thm:hiring}.
% We note it can be shown that the latter is an upper bound on the performance of any policy under the most lenient variant, and hence this implies a strong result for adaptivity gaps.

We note that solving~\eqref{lp:hiring2} for general distributions is challenging.  The argument presented here is from \citet{epstein2024selection}, who show how to solve \eqref{lp:hiring2} by reformulating it as an LP, assuming that distributions are discrete with finite support.  Concurrently, \citet{gallego2022constructive} show how to solve~\eqref{lp:hiring2} assuming that distributions are continuous, through a minimax reformulation.

All in all, in this \namecref{sec:ptk} we provided an abstract treatment of the different variants of \Cref{prob:hiring2}, which unified them into a single algorithm and result.
However, this does elude some of the richness in the structure of optimal policies for each variant.  In fact, polynomial-time approximation schemes with approximation ratios approaching 1 can be derived for some (but not all) variants; for more details, we defer to \citet{segev2021efficient}.

\section{Stochastic Knapsack} \label{sec:stocKnap}

In \Cref{sec:hiring} we studied a hiring problem in which candidates could be made offers in any order, but given offer probabilities prescribed by the LP, there was a natural order based on the weights.
In this \namecref{sec:stocKnap} we study a problem in which we need to expand the LP to prescribe an order.

\begin{problem}[Stochastic Knapsack] \label{prob:stocKnap}
There is one server for processing $n$ jobs.
Each job $i$ has a weight $W_i\ge0$ and a processing duration $D_i$ that are random and can only be realized by processing the job.
The distributions of $W_i$ and $D_i$ are known and independent across jobs.
Only one job can be processed at a time, and jobs once started cannot be cancelled.
The objective is to maximize performance, which is defined as the expected total weight of jobs completed during a time horizon of $T$.
\end{problem}

Interpreting $D_i$ as the size of a job $i$, the objective is to maximize the total weight of jobs that can fit in a knapsack of size $T$.  In the original formulation of stochastic knapsack, $W_i$ and $D_i$ were assumed to be independent from each other, which is equivalent to $W_i$ being deterministic (\citet{dean2008approximating}).  The authors used an LP that constrained only the total expected size of jobs attempted to be inserted into the knapsack:
\begin{subequations} \label{lp:stocKnap0}
\begin{align}
\max\quad & \sum_{i=1}^n \bE[W_i] y_i
\\ \mathrm{s.t.}\quad & \sum_{i=1}^n \bE[\min\{D_i,T\}] y_i \le 2T
\\ &0\le y_i\le 1 &\forall i\in[n]
\end{align}
\end{subequations}
and derived constant-factor approximation ratios by comparing to $\LP$.

In the correlated stochastic knapsack problem, any pair $(W_i,D_i)$ can follow an arbitrary given joint distribution, with these distributions still being independent across jobs $i$.  \citet{gupta2011approximation} show that LP~\eqref{lp:stocKnap0} is no longer able  to yield constant-factor approximation ratios, because it has an unbounded \textit{integrality gap}.  That is, the authors construct a family of instances for which the performance of the (trivial) optimal policy vanishes to 0, yet $\LP$ remains fixed at 1.  This means that it is not possible for an algorithm to generate a policy whose performance can maintain a constant approximation ratio relative to $\LP$ over all instances.  We present this family of instances below.

\begin{example}[\citet{gupta2011approximation}] \label{eg:timeIndexedNecessary}
Fix $T$ to be a large positive integer and let there be $n=T$ jobs.
Each job $i$ has $(W_i,D_i)$ jointly distributed to be $(1,T)$ w.p.~$1/n$ and $(0,1)$ otherwise.
That is, the job only contributes a positive weight if it takes the entire time horizon to process, which occurs with a probability of $1/n$.
Note that $\bE[\min\{D_i,T\}]=\bE[D_i]=T/n+1-1/n\le 2T/n$.
Therefore, setting $y_i=1$ for all $i\in[n]$ forms an optimal solution to LP~\eqref{lp:stocKnap0}, with objective value $n(1/n)=1$.
Meanwhile, for any actual policy, a job that started after time 1 cannot both contribute positive weight and complete in time.
Because only one job can be started at time 1, the expected total weight of jobs completed by the policy is at most $1/n$.
\end{example}

The issue with LP~\eqref{lp:stocKnap0}, as pointed out by \citet{gupta2011approximation}, is that the jobs in \Cref{eg:timeIndexedNecessary} are only relevant if they are started at time 1, which is not recognized by the LP.
\citet{gupta2011approximation} fix this issue by writing a time-indexed LP, indicating the times at which jobs are started.  This introduces the scheduling interpretation to stochastic knapsack.

We present a different time-indexed LP here, from \citet{ma2018improvements}, that yields a better approximation ratio and generalizes more directly to other stochastic knapsack problems (e.g., the variant with cancellation) and the finite-horizon Markovian bandits problem.
Although we do not elaborate on these other problems, we note that as a result their generality, the LP ends up needing the following assumption.

\begin{assumption} \label{ass:ma}
Processing durations $D_i$ are positive integers in $[T]$, and algorithm runtimes are allowed to be polynomial in $T$ (instead of needing to be polynomial in $\log T$).
\end{assumption}

Only under \Cref{ass:ma} is it permissible to write and solve the following LP whose size is polynomial in $T$:
\begin{subequations} \label{lp:stocKnap}
\begin{align}
\max\ & \sum_{i=1}^n \sum_{t=1}^T \bE[W_i\cdot\bI(D_i\le T-t+1)] y_{it}
% \label{lp:stocKnap:obj}
\\ \mathrm{s.t.\ } & \sum_{i=1}^n \sum_{u\le t} \Pr[D_i>t-u] y_{iu} \le 1 &\forall t\in[T] \label{lp:stocKnap:time}
\\ & \sum_{t=1}^T y_{it} \le 1 &\forall i\in[n] \label{lp:stocKnap:jobs}
\\ &y_{it} \ge0 &\forall i\in[n], t\in[T]
\end{align}
\end{subequations}
In this LP, $y_{it}$ is a decision variable representing the probability of starting job $i$ at time $t$.
The job completes in time if and only if $D_i\le T-t+1$, and hence starting job $i$ at time $t$ contributes an expected weight of $\bE[W_i\cdot\bI(D_i\le T-t+1)]$, noting that $W_i$ and $D_i$ can be correlated.
Meanwhile, a job $i$ started at time $u\le t$ is still processing at time $t$ if and only if $D_i>t-u$, noting that this occurs w.p.~1 if $u=t$ (because \Cref{ass:ma} says that $D_i>0$), and hence~\eqref{lp:stocKnap:time} imposes the expected number of jobs being processed at any time $t$ to not exceed 1.
Finally,~\eqref{lp:stocKnap:jobs} imposes that each job $i$ is started at most once in expectation.

\begin{lemma} \label{lem:stocKnap}
For any instance of \Cref{prob:stocKnap}, the performance of any policy is at most $\LP$.
\end{lemma}

\proof{Proof of \Cref{lem:stocKnap}.}
Fix any policy for \Cref{prob:stocKnap}.
Let $Y_{it}\in\{0,1\}$ indicate that job $i$ was started at time $t$.
On any sample path, the total weight of completed jobs is $\sum_i \sum_t W_i \bI(D_i\le T-t+1) Y_{it}$.
Meanwhile, the policy's execution satisfies $\sum_i \sum_{u\le t} \bI(D_i>t-u) Y_{iu} \le 1$ for all $t$, because at most one job can be processed at any time.
Finally, we have $\sum_t Y_{it} \le 1$ for all jobs $i$.
Now, $Y_{it}$ is independent from $W_i$ and $D_i$ for all $i\in[n]$ and $t\in[T]$, because jobs must be started without knowing their weight or duration (this argument requires the independence of $(W_i,D_i)$ across $i$).
Therefore, taking expectations allows us to deduce that setting $y_{it}=\bE[Y_{it}]$ for all $i$ and $t$ defines a feasible solution to LP~\eqref{lp:stocKnap}.  Moreover, the objective value of this feasible solution equals the performance of the policy under consideration, completing the proof.
\Halmos\endproof

We show that if we scale the probabilities $y_{it}$ in a feasible solution of LP~\eqref{lp:stocKnap} by a factor of 1/2 (which would involving idling the system more), then it is possible to start every job $i$ at every time $t$ w.p.\ exactly $y_{i,t}/2$, in an actual policy for \Cref{prob:stocKnap}.  This scaling strategy differs from the sorting strategy for randomized rounding from \Cref{sec:hiring}, and is more reminiscent of the fairness problems from \Cref{sec:selection} where we are making a promise for every job $i$ starting at every time $t$.

\begin{theorem}[\citet{ma2018improvements}] \label{thm:stocKnap}
For any instance of \Cref{prob:stocKnap}, \Cref{alg:stocKnap} starts every job $i\in[n]$ at every time $t\in[T]$ w.p.~$y_{it}/2$, and hence its performance is exactly $\LP/2$.
\end{theorem}

\begin{algorithm}[H]
\caption{for \Cref{prob:stocKnap}}\label{alg:stocKnap}
\begin{algorithmic}
\State Solve LP~\eqref{lp:stocKnap}, letting $(y_{it})_{i\in[n],t\in[T]}$ denote an optimal solution
\For{time $t=1,\ldots,T$}
\For{jobs $i\in[n]$}
\State $Y_{it}\gets0$ \Comment{indicator for starting job $i$ at time $t$; default is 0}
\State $\Free(i,t)\gets\Pr[\left(\sum_{j=1}^n \sum_{u<t} \bI(D_j>t-u) Y_{ju}=0\right)\cap\left( \sum_{u<t} Y_{iu}=0\right)]$

\Comment{probability that no job is currently being processed and $i$ has not already been started}
\EndFor
\State $i\gets$ random draw following probability vector $(\frac1{\Free(i,t)}\cdot \frac{y_{it}}{2})_{i\in[n]}$ \Comment{assume $\frac00=0$}

\Comment{$i=0$ w.p.~$1-\sum_{i=1}^n\frac1{\Free(i,t)}\cdot \frac{y_{it}}{2}$}
\If{$i>0$ \textbf{and} $\sum_{j=1}^n \sum_{u<t} \bI(D_j>t-u) Y_{ju}=0$ \textbf{and} $\sum_{u<t} Y_{iu}=0$}
\State $Y_{it}\gets 1$ \Comment{start job $i$ at time $t$}
\EndIf
\EndFor
\end{algorithmic}
\end{algorithm}

\proof{Proof of \Cref{thm:stocKnap}.}
We prove inductively over $t=1,\ldots,T$ that every job $i\in[n]$ is started at time $t$ w.p.\ exactly $y_{it}/2$.
Assuming the algorithm is valid, this follows by construction of \Cref{alg:stocKnap}, because
\begin{align*}
\bE[Y_{it}]
=\frac1{\Free(i,t)}\cdot \frac{y_{it}}{2}\Pr\left[\left(\sum_{j=1}^n \sum_{u<t} \bI(D_j>t-u) Y_{ju}=0\right)\cap\left( \sum_{u<t} Y_{iu}=0\right)\right]
=\frac{y_{it}}2.
\end{align*}
To prove that the algorithm is valid, it suffices to show
\begin{align} \label{eqn:stocKnapKey}
\sum_{i=1}^n\frac1{\Free(i,t)}\cdot \frac{y_{it}}{2} \le 1.
\end{align}
In the base case where $t=1$, we know $\Free(i,t)=1$ for all $i$, and hence~\eqref{eqn:stocKnapKey} holds by the LP constraint~\eqref{lp:stocKnap:time} for $t=1$.
For $t>1$, we derive
\begin{align*}
1-\Free(i,t)
&=\Pr\left[\left(\sum_{j=1}^n \sum_{u<t} \bI(D_j>t-u) Y_{ju}=1\right)\cup\left( \sum_{u<t} Y_{iu}=1\right)\right]
\\ &\le \Pr\left[\sum_{j=1}^n \sum_{u<t} \bI(D_j>t-u) Y_{ju}=1\right] + \Pr\left[ \sum_{u<t} Y_{iu}=1\right]
\\ &= \sum_{j=1}^n \sum_{u<t} \Pr[\bI(D_j>t-u) Y_{ju}=1] +  \sum_{u<t} \Pr[Y_{iu}=1]
\\ &= \sum_{j=1}^n \sum_{u<t} \Pr[D_j>t-u] \frac{y_{ju}}2 +  \sum_{u<t} \frac{y_{iu}}2.
\\ &\le \frac12\left(1-\sum_{j=1}^n y_{jt}\right) +  \frac12.
\end{align*}
Indeed, the second-line inequality holds by the union bound.
The third-line equality holds because at most one of the entries in $(\bI(D_j>t-u) Y_{ju})_{j\in[n],u<t}$ can be 1, and also at most one of the entries in $(Y_{iu})_{u<t}$ can be 1.
The fourth-line equality applies the induction hypothesis for time steps $u<t$.
Finally, the fifth-line inequality applies both constraints~\eqref{lp:stocKnap:time} and~\eqref{lp:stocKnap:jobs} from the LP.

Rearranging the derivation above, we get $\Free(i,t)\ge\sum_{j=1}^n\frac{y_{jt}}2$, for any $i\in[n]$.
Therefore, $\min_{i\in[n]}\Free(i,t)\ge\sum_{j=1}^n\frac{y_{jt}}2$, which implies
$$
1\ge \sum_{j=1}^n\frac{1}{\min_{i\in[n]}\Free(i,t)}\cdot\frac{y_{jt}}2\ge\sum_{j=1}^n\frac{1}{\Free(j,t)}\cdot\frac{y_{jt}}2
$$
under the convention that $\frac00=0$.  This completes the induction and the proof.
\Halmos\endproof

An issue with \Cref{alg:stocKnap}, however, is that the probabilities $\Free(i,t)$ are difficult to compute.
Indeed, they refer to events about the algorithm's state distribution at a given time $t$, which would generally have support exponential in $n$ (because the state must describe the subset of jobs not yet started).
This issue can be overcome by self-sampling, where the algorithm re-samples its past execution at each time $t$ to get estimates of $\Free(i,t)$.  It must be carefully shown that the sampling errors do not propagate, and that jobs $i$ are still started at times $t$ with probability close to $y_{it}/2$ even if both $y_{it}$ and $\Free(i,t)$ are small.
We note that tracking its own state distribution poses a general challenge for sequential randomized rounding algorithms, which need to satisfy the intricacies discussed in \Cref{rem:intricacies} to be optimal.

We defer further details to \citet{ma2018improvements}.  There it is also discussed how these techniques can be generalized to the finite-horizon Markovian bandits problem.

\section{Stochastic Matching} \label{sec:matching}

In this \namecref{sec:matching} we discuss stochastic matching in graphs, which includes many combinatorially-rich problems that exemplify the power of sequential randomized rounding.  We first prove a self-contained result for a simplified online stochastic matching model (\Cref{prob:matching}) that illustrates how the OCRSs studied in \Cref{sec:selection} can be applied.  We then describe several other models, in some cases simplified versions, to highlight the innovative LPs being rounded.

\begin{problem} \label{prob:matching}
Resources $j\in[m]$ can each be matched at most once.
Agents $i=1,\ldots,n$ arrive in order, independently "materializing" with a known probability $p_i$.
If an agent $i$ materializes, then they can be immediately matched with up to one available resource $j$, in which case a known reward ("weight") of $w_{ji}\ge 0$ is accumulated.
The objective is to maximize performance, which is defined as the expected total reward accumulated.
\end{problem}

We write the following LP for \Cref{prob:matching}, in which $x_{ji}$ represents the probability of matching resource $j$ to agent $i$:
\begin{subequations} \label{lp:matching}
\begin{align}
\max\quad & \sum_{j=1}^m \sum_{i=1}^n w_{ji} x_{ji}
\\ \mathrm{s.t.}\quad & \sum_{i=1}^n x_{ji} \le 1 &\forall j\in[m] \label{lp:matching:inv}
\\ & \sum_{j=1}^m x_{ji} \le p_i &\forall i\in[n] \label{lp:matching:dem}
\\ & x_{ji} \ge 0 &\forall j\in[m],i\in[n]
\end{align}
\end{subequations}
It can be shown similarly to the proofs of \Cref{lem:hiring,lem:stocKnap} that the optimal objective value of LP~\eqref{lp:matching} upper-bounds the performance of any policy.

\begin{theorem} \label{thm:matching}
For any instance of \Cref{prob:matching}, the performance of \Cref{alg:matching} is at least $\LP/2$.
\end{theorem}

\begin{algorithm}[H]
\caption{for \Cref{prob:matching}}\label{alg:matching}
\begin{algorithmic}
\State Solve LP~\eqref{lp:hiring}, letting $(x_{ji})_{j\in[m],i\in[n]}$ denote an optimal solution
\For{resources $j\in[m]$}
\State Run a separate version of \Cref{alg:selection} for resource $j$, with input $(x_{ji})_{i\in[n]}$ and $\gamma=1/2$, letting $(A_{ji})_{i\in[n]}$ denote the independent random bits $(A_i)_{i\in[n]}$ drawn in \Cref{alg:selection}
\EndFor
\For{agents $i=1,\ldots,n$}
\If{agent $i$ materializes}
\State Independently draw $j$ following probability vector $(\frac{x_{ji}}{p_i})_{j\in[m]}$ (with $j=0$ w.p.~$1-\sum_{j=1}^m \frac{x_{ji}}{p_i}$)
\If{resource $j$ is available \textbf{and} $A_{ji}=1$}
\State Match agent $i$ to resource $j$
\EndIf
\EndIf
\EndFor
\end{algorithmic}
\end{algorithm}

\proof{Proof of \Cref{thm:matching}.}
We first note that for each resource $j$, setting $\gamma=1/2$ in \Cref{alg:selection} is feasible, because it is no greater than the optimal guarantee of $1/(1+\sum_{i<n} x_{ji})$, which is at least 1/2 by LP constraint~\eqref{lp:matching:inv}.
We also note that $(\frac{x_{ji}}{p_i})_{j\in[m]}$ is a feasible probability vector satisfying $\sum_{j=1}^m \frac{x_{ji}}{p_i}\le 1$, by LP constraint~\eqref{lp:matching:dem}.

Having established the feasibility of \Cref{alg:matching}, we now show that every agent $i\in[n]$ is matched to every resource $j$ w.p.~$x_{ji}/2$.  By definition of \Cref{alg:selection}, every resource $j$ will be available for each agent $i$ and draw $A_{ji}=1$ w.p.~$1/2$.  Meanwhile, each agent $i$ independently materializes w.p.~$p_i$, in which case $j$ is drawn independently w.p.~$x_{ji}/p_i$.  The product of these three probabilities is exactly $x_{ji}/2$, completing the proof.
\Halmos\endproof

\Cref{thm:matching} was originally derived (for a more general agent arrival model than "materializing") in \citet{alaei2012online}, who show that the guarantee of $\LP/2$ is best-possible.  Indeed, consider a simple instance with $m=1$ resource, $n=2$ agents.  Agent 1 has $p_1=1$ and $w_1=1$.  Agent 2 has $p_2=\eps$ and $w_2=1/\eps$ for some small $\eps>0$.  In LP it is feasible to set $x_{11}=1-\eps,x_{12}=\eps$, and hence $\LP\ge(1-\eps)+(1/\eps)\eps=2-\eps$.  However, any online policy does not know whether agent 2 will materialize and hence is indifferent between accepting or rejecting agent 1, with a performance of 1.  The ratio equals $1/(2-\eps)$ which approaches 1/2 as $\eps\to0$.

To overcome this impossibility of 1/2, recent work by \citet{papadimitriou2021online} proposes a tightening of LP~\eqref{lp:matching}.  They add the following constraint to LP~\eqref{lp:matching}:
\begin{align} \label{eqn:ppsw}
x_{ji}\le p_i (1-\sum_{i'<i} x_{ji'}) &&\forall j\in[m],i\in[n]
\end{align}
which says that the probability of matching agent $i$ to resource $j$ cannot be greater than the product of the (independent) events of $i$ materializing, and $j$ not being matched to an agent $i'$ before $i$.  It is easy to check that~\eqref{eqn:ppsw} must be satisfied by any online (but not offline) policy.  Using the tightened LP, \citet{papadimitriou2021online} derive an approximation ratio strictly exceeding 1/2, which has since been improved by \citet{saberi2021greedy,braverman2022max,naor2023online}.  Approximation ratios for this problem have been recently referred to as "philosopher inequalities", with the state of the art being 0.678 (\citet{braverman2024new}).

\subsection{Online Stochastic Matching with IID Types} \label{sec:knownIID}

\Cref{thm:matching} considered a model of online stochastic matching with heterogeneous arrivals over time, but online stochastic matching was originally studied in the IID model where the arrival distribution is homogeneous over time.

\begin{problem} \label{prob:IID}
There are resources $j\in[m]$ and agent types $i\in[n]$.
Each resource $j$ can be matched at most once, and there is a known reward $w_{ji}\ge0$ for matching it to each agent type $i$.
There is a time horizon of length $T$.
Each time step $t=1,\ldots,T$, an agent arrives following a random type $i$ that is drawn IID from a known probability vector $(\lambda_i/T)_{i\in[n]}$ satisfying $\sum_{i=1}^n\lambda_i=T$.
\end{problem}

A naive LP relaxation for \Cref{prob:IID} would look like:
\begin{subequations} \label{lp:iid}
\begin{align}
\max\quad & \sum_{j=1}^m \sum_{i=1}^n w_{ji} x_{ji}
\\ \mathrm{s.t.}\quad & \sum_{i=1}^n x_{ji} \le 1 &\forall j\in[m]
\\ & \sum_{j=1}^m x_{ji} \le \lambda_i &\forall i\in[n]
\\ & x_{ji} \ge 0 &\forall j\in[m],i\in[n]
\end{align}
\end{subequations}
In this LP, $x_{ji}$ represents the expected number of times resource $j$ is matched with type $i$, noting that the expected number of arrivals of type $i$ is $\lambda_i$.

The naive LP~\eqref{lp:iid} cannot obtain a guarantee better than $1-1/e$, but a sequence of early papers (\citet{feldman2009online,manshadi2012online,jaillet2014online,brubach2016new}) show how to beat $1-1/e$ with progressively improving guarantees.
These papers need to assume either \textit{unweighted} ($w_{ji}\in\{0,1\}$ for all $i,j$) or \textit{vertex-weighted} ($w_{ji}\in\{0,w_j\}$ for all $i,j$, for some values of $(w_j)_{j\in[m]}$) graphs, or \textit{integral arrival rates} ($n=T$ and $\lambda_i=1$ for all $i\in[n]$).
Beating $1-1/e$ with neither assumption was recent accomplished by \citet{yan2024edge}, with the state-of-the-art guarantee found in \citet{qiu2023improved}.
The latter papers round the LP from \citet{jaillet2014online}, whose details are omitted.

We describe here a different "Natural LP", introduced by \citet{huang2021online}, which has led to the state of the art for unweighted and vertex-weighted graphs (without the integral arrival rates assumption).
The authors show that it suffices to consider a continuous-time version of \Cref{prob:IID}, in which the arrivals of each type $j$ follow an independent Poisson process of homogeneous rate $\lambda_j$ over [0,1].  Their Natural LP is then~\eqref{lp:iid} with the following added constraints:
\begin{align} \label{eqn:natural}
\sum_{i\in S} x_{ji} \le 1-\exp(-\sum_{i\in S}\lambda_i) &&\forall j\in[m], S\subseteq[n].
\end{align}
These constraints are valid for any online (or offline) algorithm because for any subset of types $S$, the probability of matching one of them to resource $j$ cannot exceed the probability that an agent of type $S$ exists, which occurs w.p.~$1-\exp(-\sum_{i\in S}\lambda_i)$ after the Poisson reduction.
Although~\eqref{eqn:natural} is an exponential family of constraints, \citet{huang2021online} show how to separate over them in polynomial time, which implies that the Natural LP can be solved.
The state-of-the-art guarantees for unweighted and vertex-weighted graphs are derived by combining the Natural LP with the Online Correlated Selection (OCS) technique, and we defer the details to \citet{tang2022fractional,huang2022power}.  We note that OCS was originally developed to solve open problems in adversarial settings (\citet{fahrbach2022edge,huang2024adwords}), but can be relevant even for this IID setting where the arrival probabilities and fractional solution to be rounded are given in advance, a concept recently formalized as "stochastic online correlated selection" (\citet{chen2024stochastic}).

\subsection{Stochastic Probing in Graphs} \label{sec:stocProbMatching}

Here we present a model of stochastic matching that has no "online" aspect.  Instead, it is a sequencing problem like in \Cref{sec:hiring}, where the edges in a graph can be probed in any order.

\begin{problem} \label{prob:heaven}
Let $G=(V,E)$ be a graph with vertex set $V$ and edge set $E$.
Let $\partial(v)$ denote the set of edges incident to a vertex $v$.  A subset of edges $S$ is said to be a \textit{matching} if $|S\cap\partial(v)|\le1$ for all $v\in V$, i.e.\ no two edges in $S$ are incident to the same vertex.
In this problem, each edge $e$ has a known weight $w_e$ and a probability $p_e$ with which it independently exists.
Meanwhile, each vertex has a finite patience $T_v$ upper-bounding how many times edges in $\partial(v)$ can be probed.
An edge is feasible to probe if it satisfies the patience constraint and forms a matching with the edges selected so far.
An algorithm can sequentially probe the edges in any order, but a probed edge that exists must be selected into the matching.
The objective is to maximize performance, which is defined as the expected total weight of the selected edges.
\end{problem}

\Cref{prob:heaven} is a generalization of the sequential offering problem from \Cref{sec:hiring}, and marks the third time we are seeing a problem in which the order can be freely decided.  Unlike the simpler hiring problem where there was a sorting order (\Cref{sec:hiring}), or the stochastic knapsack problem where we made the LP prescribe the order (\Cref{sec:stocKnap}), here we write an orderless LP and resort to random order to prove a guarantee:
\begin{subequations} \label{lp:probeCommit}
\begin{align}
\max\ & \sum_{e\in E} w_e p_e y_e
\\ \mathrm{s.t.\ } & \sum_{e\in\partial(v)} y_e \le T_v &\forall v\in V \label{lp:probeCommit:patience}
\\ & \sum_{e\in\partial(v)} p_e y_e \le 1 &\forall v\in V \label{lp:probeCommit:matching}
\\ & 0\le y_e\le 1 &\forall e\in E
\end{align}
\end{subequations}
In this LP, decision variable $y_e$ represents the probability of probing edge $e$.
For any vertex $v$, 
constraint~\eqref{lp:probeCommit:patience} enforces that the number of edges in $\partial(v)$ probed cannot exceed $T_v$, while
constraint~\eqref{lp:probeCommit:matching} enforces that the number of edges in $\partial(v)$ matched cannot exceed 1.

\Cref{prob:heaven} was originally formulated in \citet{chen2009approximating,bansal2012lp}, and there is a long line of work (\citet{adamczyk2015improved,baveja2018improved,brubach2021improved}) improving the approximation ratio using LP~\eqref{lp:probeCommit}.  The state-of-the-art guarantee is 0.395 (\citet{pollner2022improved}), achieved by considering the edges in a uniformly random order.

When there are no patience constraints, i.e.~$T_v=\infty$ for all $v$, \citet{gamlath2019beating} found a way to significantly tighten LP~\eqref{lp:probeCommit}.  The key idea is that for any vertex $v$, the following set of constraints validly describe any policy:
\begin{align} \label{eqn:gamlath}
\sum_{e\in S} p_e y_e \le 1-\prod_{e\in S}(1-p_e) && v\in V, S\subseteq\partial(v).
\end{align}
Indeed, the LHS of~\eqref{eqn:gamlath} represents the probability of selecting one of the edges in $S$ incident to $v$; this cannot exceed the probability of an edge in $S$ existing, which is the RHS of~\eqref{eqn:gamlath}.  This family of constraints is similar to~\eqref{eqn:natural}, and also exponentially-sized, but \citet{gamlath2019beating} establish a polynomial-time separation oracle.

Using the LP~\eqref{lp:probeCommit} with added constraints~\eqref{eqn:gamlath}, combined with the random-order allocation result from \Cref{thm:selection2}, \citet{gamlath2019beating} derive a $(1-1/e)$-approximate algorithm for bipartite graphs with infinite patience, a result that has since been improved in \citet{derakhshan2023beating}.
% reminiscent of Border's constraints (\citet{border1991implementation}
For general graphs with infinite patience, \citet{fu2021random} reduce to a notion of OCRS for graphs with random-order vertex arrivals.
They derive an 8/15-approximate algorithm for the problem on general graphs, which has since been improved in \citet{macrury2024random}.

Finally, there is a connection between \Cref{prob:heaven} and the Price-of-Information model (\citet{singla2018price}) motivated by the classical Pandora's box search problem (\citet{weitzman1978optimal}), in which one can pay a cost of $c_e$ to discover the independently random weight $W_e$ of an edge $e$.  The infinite-patience results that we discussed also hold on that model; we defer the reductions to \citet{gamlath2019beating,fu2021random}.

\subsection{Contention Resolution for Matching Polytope} \label{sec:ocrsGraphs}

Following the notation in \Cref{sec:ocrs,sec:stocProbMatching}, one can define the following OCRS problem for the matching polytope in graphs.  The agents are the edges $e\in E$, each of whom is active w.p.~$x_e$.  A subset of edges $S\subseteq E$ is feasible if and only if they form a matching, i.e.~$S$ must lie in $\cF=\{S\subseteq E: |S\cap\partial(v)|\le1 \forall v\in V\}$.  Vector $(x_e)_{e\in E}$ is restricted to lie in the matching polytope $P_\cF=\{\vx\in[0,1]^E:\sum_{e\in\partial(v)}x_e\le 1\forall v\in V\}$.

This OCRS problem was proposed by \citet{ezra2022prophet} as a means to solve the "prophet matching" problem under adversarially-ordered edge arrivals; the authors establish a selectability lower bound (algorithmic result) strictly exceeding 1/3.  Meanwhile, under randomly-ordered edge arrivals, \citet{brubach2021improved} establish a selectability lower bound of $(1-e^{-2})/2\approx0.432$.  The state of the art for both of these problems can be found in \citet{macrury2023random}.  For prophet matching under adversarially-ordered vertex arrivals, a selectability lower bound of 1/2 can be established using an argument similar to \Cref{thm:selection} (see \citet{ezra2022prophet}); for random-order vertex arrivals, the best-known selectability lower bounds are 0.535 for general graphs (\citet{macrury2024random}), $(1+e^{-2})/2\approx0.567$ for bipartite graphs with general arrivals (\citet{macrury2024random}), and $1-1/e\approx0.632$ for bipartite graphs with one side arriving before the other (\citet{ehsani2018prophet}).

\section{Conclusion and Future Directions} \label{sec:conc}

We summarize the key concepts and techniques for sequential randomized rounding (SRR) that were encountered in this tutorial.
First, the most important takeaway is perhaps the state tracking described in \Cref{rem:intricacies}, because these counter-intuitive intricacies are unique to SRR.
We illustrated these intricacies through a basic fair rationing problem (\Cref{sec:selection}), although to achieve state tracking, self-sampling can sometimes be required, as we saw in the stochastic knapsack problem (\Cref{sec:stocKnap}).  Meanwhile, we saw in the hiring problem (\Cref{sec:hiring}) an extremely simplified version of analyzing correlation, which forms the basis of a lot of advanced analyses in SRR.
Finally, SRR can be improved by tightening the LP relaxation, and we saw several examples of this in stochastic matching problems (\Cref{sec:matching}).

We now describe some other related problems in which SRR has seen recent use.
First, \citet{ashlagi2022assortment} introduce a sequential assortment optimization problem motivated by two-sided matching markets, which is reminiscent of the graph probing problem discussed in \Cref{sec:stocProbMatching}.
Algorithms for this problem are generally based on randomized rounding (\citet{torrico2020multi,rios2023platform}), with the state of the art found in \citet{housni2024two}.
Second, SRR has been fruitful at handling extensions of online decision-making models in which the length of the time horizon is stochastic.  Indeed, algorithms based on randomized rounding are used in the sequential submodular maximization problem of \citet{asadpour2023sequential}, the product ranking problem of \citet{brubach2023online} (which is just the sequential offering problem from \Cref{sec:hiring} with a random $T$), and several variants of online stochastic matching with a stochastic horizon (\citet{bai2023fluid,aouad2023nonparametric,jiang2023constant}).

We conclude by discussing three broad directions for SRR in the context of operations research applications.
First, there is a series of papers suggesting the deployment of randomized rounding algorithms in e-commerce systems (\citet{jasin2015lp,lei2022joint,ma2023order}), with the stated benefit being that the resulting policies (which are similar to the online stochastic matching policy from the beginning of \Cref{sec:matching}) are simple, fast, and parallelizable.
However, it would be important to quantify these benefits in practice, and see if the simplicity outweights any potential performance loss (see \citet{amil2022multi}).
Second, on a related note, there is often a stigma that SRR performs poorly in practice because it is designed to prove theoretical worst-case approximation ratios.
However, recent work (\citet{borodin2020experimental,ma2023fairness}) has shown this to not be the case, where simple randomized rounding policies for online matching are difficult to beat on a variety of synthetic setups.
That being said, there is still unexplained discrepancy between algorithms with the best approximation ratios and those with the best observed performances, and it is not well understood which properties of the synthetic setups or distributions may be causing this discrepancy.
Finally, randomized rounding provides a form of probabilistic fairness, which has been recently deployed by \citet{flanigan2021fair} for selecting (offline) citizens' assemblies.
It would be interesting to find other applications where probabilistic fairness is needed, which by definition would require randomized rounding, potentially in a sequential manner.
For this, we refer to two very recent papers (\citet{aminian2023markovian,banerjee2023online}) that study fairness in online stochastic settings.

% Appendix optional
%\APPENDIX{This is the Appendix Title} 
%\APPENDIX{}  % <-- This is an appendix without a title.

%\section{References}
% Make bibliography with BibTeX
% (see bibexample.tex at TutORials web site)
% Use TutORials.bst
% Note that there are no journal and proceedings abbreviations;
%   other requirements given in bibexample

\paragraph{Acknowledgements.} We thank Zhiyi Huang for a useful discussion, and anonymous reviewers for improving the Introduction of this tutorial.

\bibliographystyle{informs2014}  % put ./TutORials.bst if using locally
\bibliography{bibliography}                % put your bib file name here

\begin{thebibliography}{79}
\providecommand{\natexlab}[1]{#1}
\providecommand{\url}[1]{\texttt{#1}}
\providecommand{\urlprefix}{URL }

\bibitem[{Adamczyk et~al.(2015)Adamczyk, Grandoni, \protect\BIBand{} Mukherjee}]{adamczyk2015improved}
Adamczyk M, Grandoni F, Mukherjee J (2015) Improved approximation algorithms for stochastic matching. \emph{Algorithms-ESA 2015}, 1--12 (Springer).

\bibitem[{Alaei(2011)}]{alaei2011bayesian}
Alaei S (2011) Bayesian combinatorial auctions: Expanding single buyer mechanisms to many buyers. \emph{Proceedings of the 2011 IEEE 52nd Annual Symposium on Foundations of Computer Science}, 512--521.

\bibitem[{Alaei et~al.(2012)Alaei, Hajiaghayi, \protect\BIBand{} Liaghat}]{alaei2012online}
Alaei S, Hajiaghayi M, Liaghat V (2012) Online prophet-inequality matching with applications to ad allocation. \emph{Proceedings of the 13th ACM Conference on Electronic Commerce}, 18--35.

\bibitem[{Amil et~al.(2023)Amil, Makhdoumi, \protect\BIBand{} Wei}]{amil2022multi}
Amil A, Makhdoumi A, Wei Y (2023) Multi-item order fulfillment revisited: Lp formulation and prophet inequality. \emph{Proceedings of the 24th ACM Conference on Electronic Commerce}, 88.

\bibitem[{Aminian et~al.(2023)Aminian, Manshadi, \protect\BIBand{} Niazadeh}]{aminian2023markovian}
Aminian MR, Manshadi V, Niazadeh R (2023) Markovian search with socially aware constraints. \emph{Available at SSRN 4347447} .

\bibitem[{Aouad \protect\BIBand{} Ma(2023)}]{aouad2023nonparametric}
Aouad A, Ma W (2023) A nonparametric framework for online stochastic matching with correlated arrivals. \emph{Proceedings of the 24th ACM Conference on Electronic Commerce}, 114.

\bibitem[{Arnosti \protect\BIBand{} Ma(2023)}]{arnosti2022tight}
Arnosti N, Ma W (2023) Tight guarantees for static threshold policies in the prophet secretary problem. \emph{Operations research} 71(5):1777--1788.

\bibitem[{Asadpour et~al.(2023)Asadpour, Niazadeh, Saberi, \protect\BIBand{} Shameli}]{asadpour2023sequential}
Asadpour A, Niazadeh R, Saberi A, Shameli A (2023) Sequential submodular maximization and applications to ranking an assortment of products. \emph{Operations Research} 71(4):1154--1170.

\bibitem[{Ashlagi et~al.(2022)Ashlagi, Krishnaswamy, Makhijani, Saban, \protect\BIBand{} Shiragur}]{ashlagi2022assortment}
Ashlagi I, Krishnaswamy AK, Makhijani R, Saban D, Shiragur K (2022) Assortment planning for two-sided sequential matching markets. \emph{Operations Research} 70(5):2784--2803.

\bibitem[{Bai et~al.(2023)Bai, El~Housni, Jin, Rusmevichientong, Topaloglu, \protect\BIBand{} Williamson}]{bai2023fluid}
Bai Y, El~Housni O, Jin B, Rusmevichientong P, Topaloglu H, Williamson DP (2023) Fluid approximations for revenue management under high-variance demand. \emph{Management Science} 69(7):4016--4026.

\bibitem[{Banerjee et~al.(2023)Banerjee, Hssaine, \protect\BIBand{} Sinclair}]{banerjee2023online}
Banerjee S, Hssaine C, Sinclair SR (2023) Online fair allocation of perishable resources. \emph{ACM SIGMETRICS Performance Evaluation Review} 51(1):55--56.

\bibitem[{Bansal et~al.(2012)Bansal, Gupta, Li, Mestre, Nagarajan, \protect\BIBand{} Rudra}]{bansal2012lp}
Bansal N, Gupta A, Li J, Mestre J, Nagarajan V, Rudra A (2012) When lp is the cure for your matching woes: Improved bounds for stochastic matchings. \emph{Algorithmica} 63(4):733--762.

\bibitem[{Baveja et~al.(2018)Baveja, Chavan, Nikiforov, Srinivasan, \protect\BIBand{} Xu}]{baveja2018improved}
Baveja A, Chavan A, Nikiforov A, Srinivasan A, Xu P (2018) Improved bounds in stochastic matching and optimization. \emph{Algorithmica} 80(11):3225--3252.

\bibitem[{Blanchet et~al.(2021)Blanchet, Murthy, \protect\BIBand{} Nguyen}]{blanchet2021statistical}
Blanchet J, Murthy K, Nguyen VA (2021) Statistical analysis of wasserstein distributionally robust estimators. \emph{Tutorials in Operations Research: Emerging Optimization Methods and Modeling Techniques with Applications}, 227--254 (INFORMS).

\bibitem[{Borodin et~al.(2020)Borodin, Karavasilis, \protect\BIBand{} Pankratov}]{borodin2020experimental}
Borodin A, Karavasilis C, Pankratov D (2020) An experimental study of algorithms for online bipartite matching. \emph{Journal of Experimental Algorithmics (JEA)} 25:1--37.

\bibitem[{Braverman et~al.(2022)Braverman, Derakhshan, \protect\BIBand{} Molina~Lovett}]{braverman2022max}
Braverman M, Derakhshan M, Molina~Lovett A (2022) Max-weight online stochastic matching: Improved approximations against the online benchmark. \emph{Proceedings of the 23rd ACM Conference on Economics and Computation}, 967--985.

\bibitem[{Braverman et~al.(2024)Braverman, Derakhshan, Pollner, Saberi, \protect\BIBand{} Wajc}]{braverman2024new}
Braverman M, Derakhshan M, Pollner T, Saberi A, Wajc D (2024) New philosopher inequalities for online bayesian matching, via pivotal sampling. \emph{arXiv preprint arXiv:2407.15285} .

\bibitem[{Brubach et~al.(2021)Brubach, Grammel, Ma, \protect\BIBand{} Srinivasan}]{brubach2021improved}
Brubach B, Grammel N, Ma W, Srinivasan A (2021) Improved guarantees for offline stochastic matching via new ordered contention resolution schemes. \emph{Advances in Neural Information Processing Systems} 34:27184--27195.

\bibitem[{Brubach et~al.(2023)Brubach, Grammel, Ma, \protect\BIBand{} Srinivasan}]{brubach2023online}
Brubach B, Grammel N, Ma W, Srinivasan A (2023) Online matching frameworks under stochastic rewards, product ranking, and unknown patience. \emph{Operations Research} .

\bibitem[{Brubach et~al.(2016)Brubach, Sankararaman, Srinivasan, \protect\BIBand{} Xu}]{brubach2016new}
Brubach B, Sankararaman KA, Srinivasan A, Xu P (2016) New algorithms, better bounds, and a novel model for online stochastic matching. \emph{24th Annual European Symposium on Algorithms (ESA 2016)} (Schloss Dagstuhl-Leibniz-Zentrum fuer Informatik).

\bibitem[{Chawla \protect\BIBand{} Sivan(2014)}]{chawla2014bayesian}
Chawla S, Sivan B (2014) Bayesian algorithmic mechanism design. \emph{ACM SIGecom Exchanges} 13(1):5--49.

\bibitem[{Chekuri et~al.(2014)Chekuri, Vondrak, \protect\BIBand{} Zenklusen}]{chekuri2014submodular}
Chekuri C, Vondrak J, Zenklusen R (2014) Submodular function maximization via the multilinear relaxation and contention resolution schemes. \emph{SIAM Journal on Computing} 43(6):1831--1879.

\bibitem[{Chen et~al.(2009)Chen, Immorlica, Karlin, Mahdian, \protect\BIBand{} Rudra}]{chen2009approximating}
Chen N, Immorlica N, Karlin AR, Mahdian M, Rudra A (2009) Approximating matches made in heaven. \emph{Automata, Languages and Programming: 36th International Colloquium, ICALP 2009, Rhodes, Greece, July 5-12, 2009, Proceedings, Part I 36}, 266--278 (Springer).

\bibitem[{Chen et~al.(2024)Chen, Huang, \protect\BIBand{} Sun}]{chen2024stochastic}
Chen Z, Huang Z, Sun E (2024) Stochastic online correlated selection. \emph{arXiv preprint arXiv:2408.12524} .

\bibitem[{Correa et~al.(2019)Correa, Foncea, Hoeksma, Oosterwijk, \protect\BIBand{} Vredeveld}]{correa2019recent}
Correa J, Foncea P, Hoeksma R, Oosterwijk T, Vredeveld T (2019) Recent developments in prophet inequalities. \emph{ACM SIGecom Exchanges} 17(1):61--70.

\bibitem[{Dean et~al.(2008)Dean, Goemans, \protect\BIBand{} Vondr{\'a}k}]{dean2008approximating}
Dean BC, Goemans MX, Vondr{\'a}k J (2008) Approximating the stochastic knapsack problem: The benefit of adaptivity. \emph{Mathematics of Operations Research} 33(4):945--964.

\bibitem[{Derakhshan \protect\BIBand{} Farhadi(2023)}]{derakhshan2023beating}
Derakhshan M, Farhadi A (2023) Beating (1-1/e)-approximation for weighted stochastic matching. \emph{Proceedings of the 2023 Annual ACM-SIAM Symposium on Discrete Algorithms (SODA)}, 1931--1961 (SIAM).

\bibitem[{Dughmi(2019)}]{dughmi2019outer}
Dughmi S (2019) The outer limits of contention resolution on matroids and connections to the secretary problem. \emph{arXiv preprint arXiv:1909.04268} .

\bibitem[{Ehsani et~al.(2018)Ehsani, Hajiaghayi, Kesselheim, \protect\BIBand{} Singla}]{ehsani2018prophet}
Ehsani S, Hajiaghayi M, Kesselheim T, Singla S (2018) Prophet secretary for combinatorial auctions and matroids. \emph{Proceedings of the twenty-ninth annual acm-siam symposium on discrete algorithms}, 700--714 (SIAM).

\bibitem[{Epstein \protect\BIBand{} Ma(2024)}]{epstein2024selection}
Epstein B, Ma W (2024) Selection and ordering policies for hiring pipelines via linear programming. \emph{Operations Research} .

\bibitem[{Ezra et~al.(2022)Ezra, Feldman, Gravin, \protect\BIBand{} Tang}]{ezra2022prophet}
Ezra T, Feldman M, Gravin N, Tang ZG (2022) Prophet matching with general arrivals. \emph{Mathematics of Operations Research} 47(2):878--898.

\bibitem[{Fahrbach et~al.(2022)Fahrbach, Huang, Tao, \protect\BIBand{} Zadimoghaddam}]{fahrbach2022edge}
Fahrbach M, Huang Z, Tao R, Zadimoghaddam M (2022) Edge-weighted online bipartite matching. \emph{Journal of the ACM} 69(6):1--35.

\bibitem[{Feldman et~al.(2009)Feldman, Mehta, Mirrokni, \protect\BIBand{} Muthukrishnan}]{feldman2009online}
Feldman J, Mehta A, Mirrokni V, Muthukrishnan S (2009) Online stochastic matching: Beating 1-1/e. \emph{2009 50th Annual IEEE Symposium on Foundations of Computer Science}, 117--126 (IEEE).

\bibitem[{Feldman et~al.(2021)Feldman, Svensson, \protect\BIBand{} Zenklusen}]{feldman2021online}
Feldman M, Svensson O, Zenklusen R (2021) Online contention resolution schemes with applications to bayesian selection problems. \emph{SIAM Journal on Computing} 50(2):255--300.

\bibitem[{Flanigan et~al.(2021)Flanigan, G{\"o}lz, Gupta, Hennig, \protect\BIBand{} Procaccia}]{flanigan2021fair}
Flanigan B, G{\"o}lz P, Gupta A, Hennig B, Procaccia AD (2021) Fair algorithms for selecting citizens’ assemblies. \emph{Nature} 596(7873):548--552.

\bibitem[{Fu et~al.(2021)Fu, Tang, Wu, Wu, \protect\BIBand{} Zhang}]{fu2021random}
Fu H, Tang ZG, Wu H, Wu J, Zhang Q (2021) Random order vertex arrival contention resolution schemes for matching, with applications. \emph{48th International Colloquium on Automata, Languages, and Programming (ICALP 2021)} (Schloss-Dagstuhl-Leibniz Zentrum f{\"u}r Informatik).

\bibitem[{Gallego \protect\BIBand{} Segev(2022)}]{gallego2022constructive}
Gallego G, Segev D (2022) A constructive prophet inequality approach to the adaptive probemax problem. \emph{arXiv preprint arXiv:2210.07556} .

\bibitem[{Gamlath et~al.(2019)Gamlath, Kale, \protect\BIBand{} Svensson}]{gamlath2019beating}
Gamlath B, Kale S, Svensson O (2019) Beating greedy for stochastic bipartite matching. \emph{Proceedings of the Thirtieth Annual ACM-SIAM Symposium on Discrete Algorithms}, 2841--2854 (SIAM).

\bibitem[{Gandhi et~al.(2006)Gandhi, Khuller, Parthasarathy, \protect\BIBand{} Srinivasan}]{gandhi2006dependent}
Gandhi R, Khuller S, Parthasarathy S, Srinivasan A (2006) Dependent rounding and its applications to approximation algorithms. \emph{Journal of the ACM (JACM)} 53(3):324--360.

\bibitem[{Gupta et~al.(2011)Gupta, Krishnaswamy, Molinaro, \protect\BIBand{} Ravi}]{gupta2011approximation}
Gupta A, Krishnaswamy R, Molinaro M, Ravi R (2011) Approximation algorithms for correlated knapsacks and non-martingale bandits. \emph{2011 IEEE 52nd Annual Symposium on Foundations of Computer Science}, 827--836 (IEEE).

\bibitem[{Housni et~al.(2024)Housni, Torrico, \protect\BIBand{} Hennebelle}]{housni2024two}
Housni OE, Torrico A, Hennebelle U (2024) Two-sided assortment optimization: Adaptivity gaps and approximation algorithms. \emph{arXiv preprint arXiv:2403.08929} .

\bibitem[{Huang \protect\BIBand{} Shu(2021)}]{huang2021online}
Huang Z, Shu X (2021) Online stochastic matching, poisson arrivals, and the natural linear program. \emph{Proceedings of the 53rd Annual ACM SIGACT Symposium on Theory of Computing}, 682--693.

\bibitem[{Huang et~al.(2022)Huang, Shu, \protect\BIBand{} Yan}]{huang2022power}
Huang Z, Shu X, Yan S (2022) The power of multiple choices in online stochastic matching. \emph{Proceedings of the 54th Annual ACM SIGACT Symposium on Theory of Computing}, 91--103.

\bibitem[{Huang et~al.(2024{\natexlab{a}})Huang, Tang, \protect\BIBand{} Wajc}]{huang2024online}
Huang Z, Tang ZG, Wajc D (2024{\natexlab{a}}) Online matching: A brief survey. \emph{SIGECOM Exchanges} .

\bibitem[{Huang \protect\BIBand{} Tr{\"o}bst(2023)}]{huang2023applications}
Huang Z, Tr{\"o}bst T (2023) Applications of online matching. \emph{Online and Matching-Based Market Design} 109.

\bibitem[{Huang et~al.(2024{\natexlab{b}})Huang, Zhang, \protect\BIBand{} Zhang}]{huang2024adwords}
Huang Z, Zhang Q, Zhang Y (2024{\natexlab{b}}) Adwords in a panorama. \emph{SIAM Journal on Computing} 53(3):701--763.

\bibitem[{Jaillet \protect\BIBand{} Lu(2014)}]{jaillet2014online}
Jaillet P, Lu X (2014) Online stochastic matching: New algorithms with better bounds. \emph{Mathematics of Operations Research} 39(3):624--646.

\bibitem[{Jasin \protect\BIBand{} Sinha(2015)}]{jasin2015lp}
Jasin S, Sinha A (2015) An lp-based correlated rounding scheme for multi-item ecommerce order fulfillment. \emph{Operations Research} 63(6):1336--1351.

\bibitem[{Jiang(2023)}]{jiang2023constant}
Jiang J (2023) Constant approximation for network revenue management with markovian-correlated customer arrivals. \emph{arXiv preprint arXiv:2305.05829} .

\bibitem[{Jiang et~al.(2022)Jiang, Ma, \protect\BIBand{} Zhang}]{jiang2022tight}
Jiang J, Ma W, Zhang J (2022) Tight guarantees for multi-unit prophet inequalities and online stochastic knapsack. \emph{Proceedings of the 2022 Annual ACM-SIAM Symposium on Discrete Algorithms (SODA)}, 1221--1246 (SIAM).

\bibitem[{Kleinberg \protect\BIBand{} Weinberg(2012)}]{kleinberg2012matroid}
Kleinberg R, Weinberg SM (2012) Matroid prophet inequalities. \emph{Proceedings of the forty-fourth annual ACM symposium on Theory of computing}, 123--136.

\bibitem[{Kuhn et~al.(2019)Kuhn, Esfahani, Nguyen, \protect\BIBand{} Shafieezadeh-Abadeh}]{kuhn2019wasserstein}
Kuhn D, Esfahani PM, Nguyen VA, Shafieezadeh-Abadeh S (2019) Wasserstein distributionally robust optimization: Theory and applications in machine learning. \emph{Operations research \& management science in the age of analytics}, 130--166 (Informs).

\bibitem[{Lee \protect\BIBand{} Singla(2018)}]{lee2018optimal}
Lee E, Singla S (2018) Optimal online contention resolution schemes via ex-ante prophet inequalities. \emph{26th European Symposium on Algorithms, ESA 2018}, . (Schloss Dagstuhl-Leibniz-Zentrum fur Informatik GmbH, Dagstuhl Publishing).

\bibitem[{Lei et~al.(2022)Lei, Jasin, Uichanco, \protect\BIBand{} Vakhutinsky}]{lei2022joint}
Lei Y, Jasin S, Uichanco J, Vakhutinsky A (2022) Joint product framing (display, ranking, pricing) and order fulfillment under the multinomial logit model for e-commerce retailers. \emph{Manufacturing \& Service Operations Management} 24(3):1529--1546.

\bibitem[{Lucier(2017)}]{lucier2017economic}
Lucier B (2017) An economic view of prophet inequalities. \emph{ACM SIGecom Exchanges} 16(1):24--47.

\bibitem[{Ma(2018)}]{ma2018improvements}
Ma W (2018) Improvements and generalizations of stochastic knapsack and markovian bandits approximation algorithms. \emph{Mathematics of Operations Research} 43(3):789--812.

\bibitem[{Ma(2023)}]{ma2023order}
Ma W (2023) Order-optimal correlated rounding for fulfilling multi-item e-commerce orders. \emph{Manufacturing \& Service Operations Management} 25(4):1324--1337.

\bibitem[{Ma et~al.(2023)Ma, Xu, \protect\BIBand{} Xu}]{ma2023fairness}
Ma W, Xu P, Xu Y (2023) Fairness maximization among offline agents in online-matching markets. \emph{ACM Transactions on Economics and Computation} 10(4):1--27.

\bibitem[{MacRury \protect\BIBand{} Ma(2023)}]{macrury2024random}
MacRury C, Ma W (2023) Random-order contention resolution via continuous induction: Tightness for bipartite matching under vertex arrivals. \emph{arXiv preprint arXiv:2310.10101} .

\bibitem[{MacRury et~al.(2023)MacRury, Ma, \protect\BIBand{} Grammel}]{macrury2023random}
MacRury C, Ma W, Grammel N (2023) On (random-order) online contention resolution schemes for the matching polytope of (bipartite) graphs. \emph{Proceedings of the 2023 Annual ACM-SIAM Symposium on Discrete Algorithms (SODA)}, 1995--2014 (SIAM).

\bibitem[{Manshadi et~al.(2012)Manshadi, Gharan, \protect\BIBand{} Saberi}]{manshadi2012online}
Manshadi VH, Gharan SO, Saberi A (2012) Online stochastic matching: Online actions based on offline statistics. \emph{Mathematics of Operations Research} 37(4):559--573.

\bibitem[{Mehta et~al.(2013)}]{mehta2013online}
Mehta A, et~al. (2013) Online matching and ad allocation. \emph{Foundations and Trends{\textregistered} in Theoretical Computer Science} 8(4):265--368.

\bibitem[{Naor et~al.(2023)Naor, Srinivasan, \protect\BIBand{} Wajc}]{naor2023online}
Naor JS, Srinivasan A, Wajc D (2023) Online dependent rounding schemes. \emph{arXiv preprint arXiv:2301.08680} .

\bibitem[{Papadimitriou et~al.(2021)Papadimitriou, Pollner, Saberi, \protect\BIBand{} Wajc}]{papadimitriou2021online}
Papadimitriou C, Pollner T, Saberi A, Wajc D (2021) Online stochastic max-weight bipartite matching: Beyond prophet inequalities. \emph{Proceedings of the 22nd ACM Conference on Economics and Computation}, 763--764.

\bibitem[{Pollner et~al.(2023)Pollner, Roghani, Saberi, \protect\BIBand{} Wajc}]{pollner2022improved}
Pollner T, Roghani M, Saberi A, Wajc D (2023) Improved online contention resolution for matchings and applications to the gig economy. \emph{Mathematics of Operations Research} .

\bibitem[{Purohit et~al.(2019)Purohit, Gollapudi, \protect\BIBand{} Raghavan}]{purohit2019hiring}
Purohit M, Gollapudi S, Raghavan M (2019) Hiring under uncertainty. \emph{International Conference on Machine Learning}, 5181--5189 (PMLR).

\bibitem[{Qiu \protect\BIBand{} Singla(2022)}]{qiu2022submodular}
Qiu F, Singla S (2022) Submodular dominance and applications. \emph{arXiv preprint arXiv:2207.04957} .

\bibitem[{Qiu et~al.(2023)Qiu, Feng, Zhou, \protect\BIBand{} Wu}]{qiu2023improved}
Qiu G, Feng Y, Zhou S, Wu X (2023) Improved competitive ratio for edge-weighted online stochastic matching. \emph{International Conference on Web and Internet Economics}, 527--544 (Springer).

\bibitem[{Rios \protect\BIBand{} Torrico(2023)}]{rios2023platform}
Rios I, Torrico A (2023) Platform design in matching markets: A two-sided assortment optimization approach. \emph{arXiv preprint arXiv:2308.02584} .

\bibitem[{Saberi \protect\BIBand{} Wajc(2021)}]{saberi2021greedy}
Saberi A, Wajc D (2021) The greedy algorithm is not optimal for on-line edge coloring. \emph{48th International Colloquium on Automata, Languages, and Programming (ICALP 2021)} (Schloss Dagstuhl-Leibniz-Zentrum f{\"u}r Informatik).

\bibitem[{Segev \protect\BIBand{} Singla(2021)}]{segev2021efficient}
Segev D, Singla S (2021) Efficient approximation schemes for stochastic probing and prophet problems. \emph{Proceedings of the 22nd ACM Conference on Economics and Computation}, 793--794.

\bibitem[{Singla(2018)}]{singla2018price}
Singla S (2018) The price of information in combinatorial optimization. \emph{Proceedings of the twenty-ninth annual ACM-SIAM symposium on discrete algorithms}, 2523--2532 (SIAM).

\bibitem[{Tang et~al.(2022)Tang, Wu, \protect\BIBand{} Wu}]{tang2022fractional}
Tang ZG, Wu J, Wu H (2022) (fractional) online stochastic matching via fine-grained offline statistics. \emph{Proceedings of the 54th Annual ACM SIGACT Symposium on Theory of Computing}, 77--90.

\bibitem[{Torrico et~al.(2020)Torrico, Carvalho, \protect\BIBand{} Lodi}]{torrico2020multi}
Torrico A, Carvalho M, Lodi A (2020) Multi-agent assortment optimization in sequential matching markets. \emph{arXiv preprint arXiv:2006.04313} .

\bibitem[{Vazirani(2001)}]{vazirani2001approximation}
Vazirani VV (2001) \emph{Approximation algorithms}, volume~1 (Springer).

\bibitem[{Weitzman(1978)}]{weitzman1978optimal}
Weitzman M (1978) \emph{Optimal search for the best alternative}, volume~78 (Department of Energy).

\bibitem[{Williamson \protect\BIBand{} Shmoys(2011)}]{williamson2011design}
Williamson DP, Shmoys DB (2011) \emph{The design of approximation algorithms} (Cambridge university press).

\bibitem[{Yan(2011)}]{yan2011mechanism}
Yan Q (2011) Mechanism design via correlation gap. \emph{Proceedings of the twenty-second annual ACM-SIAM symposium on Discrete Algorithms}, 710--719 (SIAM).

\bibitem[{Yan(2024)}]{yan2024edge}
Yan S (2024) Edge-weighted online stochastic matching: Beating. \emph{Proceedings of the 2024 Annual ACM-SIAM Symposium on Discrete Algorithms (SODA)}, 4631--4640 (SIAM).

\end{thebibliography}

%%%%%%%%%%%%%%%%
\end{document}